\documentclass[preprint,aps,prc]{revtex4-1}
\usepackage{graphicx}
\usepackage{epsfig}
\def\Vec#1{\mbox{\boldmath $#1$}}

\begin{document}

\title{
$\alpha+\alpha+t$ cluster structures and Hoyle-analogue states in $^{11}$B
}

\author{Taiichi Yamada}
\affiliation{Laboratory of Physics, Kanto Gakuin University, Yokohama 236-8501, Japan
}

\author{Yasuro Funaki}
\affiliation{Institute of Physics, University of Tsukuba, Tsukuba 305-8571, Japan
}

\date{\today}

\begin{abstract}
The structure of $3/2^{-}$ and $1/2^{+}$ states in $^{11}$B is investigated 
 with an $\alpha+\alpha+t$ orthogonality condition model (OCM) based on the Gaussian expansion method.
Full levels up to the $3/2^{-}_{3}$ and $1/2^{+}_2$ states around the $\alpha+\alpha+t$ threshold ($E_x$=11.1 MeV)
 are reproduced consistently with the experimental energy levels.
It is shown that the $3/2_{3}^{-}$ state located around the $^{7}$Li+$\alpha$ threshold has
 an $\alpha+\alpha+t$ cluster structure, whereas the $3/2_{1}^{-}$ and $3/2_{2}^{-}$ states 
 have a shell-model-like compact structure.
We found that the $3/2_{3}^{-}$ state does not possess an $\alpha$-condensate-like nature
 similar to the $0^{+}_{2}$ state of $^{12}$C (Hoyle state) which has a dilute $3\alpha$-condensate
 structure described by a $(0S_{\alpha})^3$ configuration with about $70$\% probability, 
 although the monopole transition strength of the former is as large as that of the latter.
We discuss the reasons why the $3/2_{3}^{-}$ state does not have the condensate character.
On the other hand, the $1/2^{+}_{1}$ state just below the $^{7}$Li+$\alpha$ threshold 
 has a cluster structure which can be interpreted as a parity-doublet partner of the $3/2^{-}_3$ state. 
We indicate that the $12.56$-MeV state ($J^{\pi}=1/2^{+}_{2}$) just above the $\alpha+\alpha+t$ 
 threshold observed in the $^7$Li($^{7}$Li,$^{11}$B$^*$)$t$ reaction etc.~is of the dilute-cluster-gas-like,
 and is a strong candidate for the Hoyle-analogue state which has a configuration of 
 $(0S_{\alpha})^{2}(0S_{t})$ with about $65$\% probability from the analyses of the single-cluster motions in $^{11}$B. 
The structure property of the $1/2^{+}$ resonant state is analyzed with the complex scaling method.  
\\\\  
%{PACS numbers: 21.10.Dr, 21.10.Gv, 21.60.Gx, 03.75.Hh}\\
\end{abstract}

\maketitle

%%%%%%%%%%%%%%%%%%%%%%%%%%%%%%%%%%%%%%%%%%%%%%%%%%%%
\section{Introduction}\label{sec:intro}
%%%%%%%%%%%%%%%%%%%%%%%%%%%%%%%%%%%%%%%%%%%%%%%%%%%%

Cluster picture as well as mean-field picture is important to understand the structure of
 light nuclei~\cite{wildermuth77,ikeda80}.
The Hoyle state (the $0^+_2$ state at $E_{x}=7.65$ MeV in $^{12}$C) is a typical cluster state
 with the $3\alpha$ cluster structure~\cite{uegaki77,kamimura77}, which appears by only $0.38$ MeV above
 the $3\alpha$ threshold and is characterized by the large monopole transition rate sharing
 about 16~\% of the energy-weighted sum rule~\cite{yamada08_monopole}.
Recently the Hoyle state has been reinvestigated
 from the viewpoint of the $\alpha$ condensation~\cite{tohsaki01}.
The definitions and occurrences of the nuclear $\alpha$-particle condensation are discussed
 in detail in Ref.~\cite{funaki09}. 
Many theoretical works have showed that the Hoyle state has a $3\alpha$-condensate-like
 structure~\cite{tohsaki01,funaki03,yamada05,matsumura04,schuck07,funaki08,funaki09,chernykh07},
 in which the $3\alpha$ particles occupy an identical $0S$-orbit with $70~\%$
 probability~\cite{yamada05,matsumura04}, forming a dilute $\alpha$-gas-like configuration with
 $(0S_{\alpha})^3$, i.e.~the product state of three $\alpha$'s.

The structure study of $^{16}$O has recently made a great advance.
The six lowest $0^{+}$ states of $^{16}$O have very nicely been reproduced up to
 about $15$~MeV excitation energy, including the ground state, 
 with the $4\alpha$ orthogonality condition model (OCM)~\cite{funaki08_16O}.
The OCM is a semi-microscopic cluster model, which is an approximation of RGM (resonating group method)
 and is extensively described in Ref.~\cite{saitoh68}.
Many successful applications of OCM are reported in Ref.~\cite{ikeda80}.
The $4\alpha$ OCM calculation showed that the $0^+_6$ state around the $4\alpha$ threshold is a strong candidate
 of the $4\alpha$-particle condensate state, having a large $\alpha$ condensate fraction of $60$~\%
 for the $\alpha$-gas-like configuration $(0S_{\alpha})^{4}$ as well as 
 a large component of $\alpha+^{12}$C$(0_2^+)$ configuration.
The $0^{+}_{6}$ state could be called the Hoyle-analogue state. 
The Hoyle-analogue states heavier than $^{16}$O are predicted to exist
 around their $\alpha$-cluster disintegrated thresholds in self-conjugate $A=4n$
 nuclei~\cite{tohsaki01,tohsaki04,yamada04}.  

Besides the $4n$ nuclei, one can also expect cluster-gas states composed of alpha and triton
 clusters (including valence neutrons etc.) around their cluster disintegrated thresholds in $A \not= 4n$ nuclei, 
 in which all clusters are in their respective $0S$ orbits, similar to the Hoyle state with $(0S_{\alpha})^3$.
The states, thus, can be called {\it Hoyle-analogue} in the non-self-conjugated nuclei.
It is an intriguing subject to investigate whether or not the Hoyle-analogue states
 exist in $A \not= 4n$ nuclei, for example, $^{11}$B, composed of $2\alpha$ and $t$ clusters. 

The cluster structure in $^{11}$B was studied about 30 years ago with the $\alpha+\alpha+t$ OCM
 using the harmonic oscillator basis by Nishioka et al.~\cite{nishioka79,furutani80}.
They found that 1)~the $3/2^-_3$ state ($E_{x}=8.6$~MeV)
 just below the $^7{\rm Li}+\alpha$ threshold has a well-developed molecular-like structure
 of $^7$Li(g.s)+$\alpha$, where $^7$Li(g.s) denotes the ground state of $^{7}$Li
 with the $\alpha$+$t$ cluster structure, and 2)~the $3/2^{-}_{1}$~(g.s) and $3/2^{-}_{2}$ ($E_{x}=5.0$~MeV)
 states have shell-model-like compact structures. 
The model space adopted in Ref.~\cite{nishioka79}, however, was not sufficient to account simultaneously 
 for the $^{7}$Li+$\alpha$ cluster configuration as well as the $\alpha+\alpha+t$ gas-like configuration.
In addition, limited experimental data in those days caused difficulties in giving the definite conclusion 
 that the $3/2^-_3$ state has an $\alpha+\alpha+t$ cluster structure.

Recently Kawabata and his collaborators have investigated the excited states  
 of $^{11}$B by using the ${^{11}{\rm B}}(d, d')$ reaction~\cite{kawabata07}.
They eventually concluded that the $3/2^{-}_{3}$ state at $E_x$ = 8.56 MeV has an 
 $\alpha$-cluster structure.  
Among many reasons for this conclusion, one is a large isoscalar monopole 
 transition rate for the $3/2^{-}_{3}$ state, $B{\rm (E0:IS)}=96\pm16$~fm$^{4}$, 
 which is of the similar value to that for the Hoyle state in $^{12}$C,
 $B{\rm (E0:IS)}=120\pm9$~fm$^{4}$.
A close relation between large isoscalar monopole strengths and underlying cluster structures
 in excited states were discussed by the present authors et al.~(see Ref.~\cite{yamada08_monopole}). 
According to the literature, the largeness of the isoscalar monopole transition rate
 for $3/2^{-}_{3}$ originates from the fact that the state has a cluster structure. 

Another reason why the $3/2^{-}_{3}$ state has the cluster structure is that
 the AMD (antisymmetrized molecular dynamics)
 calculation~\cite{kawabata07,enyo07} has reproduced the large monopole transition rate
 for the $3/2^{-}_{3}$ state and has succeeded in assigning an $\alpha+\alpha+t$ cluster structure to this state. 
Comparing the density distribution of the $3/2^{-}_{3}$ state (calculated by AMD)
 with that of the Hoyle state (together with the analysis of the expectation values of
 the harmonic oscillator quanta in their wave functions),  
 the AMD calculation~\cite{enyo07} claimed that the $3/2^{-}_{3}$ state
 has a clustering feature similar to that of the Hoyle state.
However, they did not study single-cluster properties such as single-cluster orbits
 and their occupation probabilities in the $3/2^{-}_{3}$ state.
Their quantities are independent on the information of the density distribution 
 and the expectation values of the harmonic oscillator quanta in the wave function.
Thus, it is very important to study the single-cluster properties
 in order to judge whether the $3/2^{-}_{3}$ state possesses the similarity 
 to the Hoyle state with the $(0S_{\alpha})^3$-like structure.    
As discussed in Refs.~\cite{yamada05,matsumura04,suzuki02,yamada08_density}, the single-cluster motions
 such as $\alpha$ ($t$) cluster orbits and their occupation probabilities
 in a nuclear state can be investigated by solving the eigenvalue equation
 of the single-cluster density matrix $\rho(\Vec{r},\Vec{r}')$ derived from 
 the microscopic and/or semi-microscopic wave functions.
 
The purpose of the present paper is to study the structure of $3/2^-$ and $1/2^+$ states
 in $^{11}$B up to around the $\alpha+\alpha+t$ threshold.
Here we take the $\alpha+\alpha+t$ OCM with the Gaussian expansion method (GEM),
 the model space of which is large enough to cover the $\alpha+\alpha+t$ gas,
 the $^{7}$Li+$\alpha$ cluster, as well as the shell-model configurations. 
Combining OCM and GEM provides a powerful method to study the structure of light
 nuclei~\cite{yamada05,funaki08,yamada08} as well as light hypernuclei~\cite{hiyama97,hiyama09},
 because the Pauli-blocking effect among the clusters is properly taken into account
 by OCM and GEM covers an approximately complete model space~\cite{kamimura88,hiyama03}.
This framework can treat precisely a strong parity dependence of the $\alpha-t$
 potential:~The negative-parity potential is attractive enough to make bound states
 ($3/2^{-}_{1}$ and $1/2^{-}_{1}$) and resonant states ($7/2^{-}_{1}$ and $5/2^{-}_{1}$)
 of $^{7}$Li, wheres the positive-parity potential is not.
This parity dependence should play an important role in producing the cluster states of $^{11}$B, 
 as will be discussed later.
The single-cluster properties such as single-$\alpha$-particle (single-$t$-particle) orbits and
 occupation probabilities are investigated to judge whether or not the $3/2^{-}_3$ state has
 a Hoyle-analogue structure by solving the eigenvalue equation of
 the single cluster density matrix $\rho(\Vec{r},\Vec{r}')$ derived from
 the total wave function of $^{11}$B.
The resonance structure of $1/2^{+}$ states is also studied with
 the complex scaling method (CSM), since they appear around the $\alpha+\alpha+t$ threshold
 with $E_{x}=11.1$ MeV.
 
It is also interesting to explore the Hoyle-analogue state
 with the $\alpha+\alpha+t$ structure in $1/2^{+}$ states,
 because one can conjecture an appearance of the product state,
 $(0S_{\alpha})^2(0S_{t})$, in $J^{\pi}=1/2^{+}$ of $^{11}$B
 ($1/2$ comes from the spin of triton) from the similarity of $(0S_{\alpha})^3$
 in the Hoyle state.  
The structure of the positive-parity states (isospin $T=1/2$) in $^{11}$B so far has not been
 discussed well in AMD~\cite{enyo07} and non-core shell model~\cite{NCSM09}. 
Recently several positive-parity states of $^{11}$B with $T=1/2$ are observed
 with the $^7$Li($^9$Be,$\alpha^7{\rm Li}$)$^5$He reaction~\cite{soic04},
 the $^7$Li($^{7}$Li,$^{11}$B$^*$)$t$ reaction~\cite{curtis05}, and 
 the decay-particle measurements of $\alpha$+$^7$Li and $t$+$^8$Be from
 excited $^{11}$B$^*$ states~\cite{charity08}.
An interesting result in their experiments is as follows:~The $1/2^+(3/2^+)$ state
 at $E_{x}=12.56$ MeV
 which was identified so far as isospin $T=3/2$ state~\cite{ajzenberg86},
 is observed through the $\alpha$+$^7$Li decay channel~\cite{soic04,curtis05,charity08}.
They concluded that the $12.56$-MeV state has isospin $T=1/2$ or 
 alternatively it is observed at a similar energy to that with $T=3/2$.   
Therefore, it is interesting to investigate theoretically the structure
 of the 12.56-MeV state.
 
The present paper is organized as follows.  
In Sec.~\ref{sec:formulation} we formulate the $\alpha+\alpha+t$ OCM with GEM as well as CSM.
The results and discussion are devoted in Sec.~\ref{sec:results_discussions}.
Finally we present the summary in Sec.~\ref{sec:summary}.

%%%%%%%%%%%% 
%  Fig. 1  %
%%%%%%%%%%%%
\begin{figure}[t]
\begin{center}
\epsfig{file=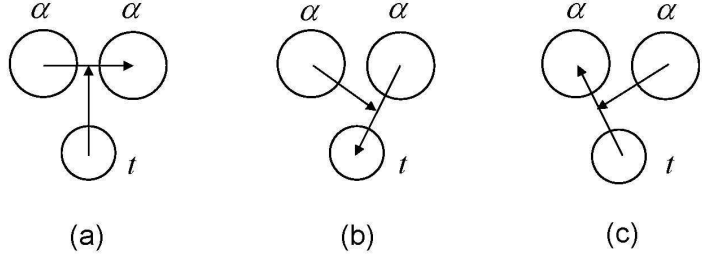,width=0.6\hsize}
\caption{
Three Jacobi-coordinate systems in the $\alpha+\alpha+t$ model.
}
\label{fig:1}
\end{center}
\end{figure}

%\newpage
%%%%%%%%%%%%%%%%%%%%%%%%%%%%%%%%%%%%%%%%%%%%%%%%%%%%
\section{Formulation}\label{sec:formulation}
%%%%%%%%%%%%%%%%%%%%%%%%%%%%%%%%%%%%%%%%%%%%%%%%%%%%

In this section, we present the formulation of the $\alpha+\alpha+t$ OCM
 with GEM for $^{11}$B, and give a brief formulation of CSM with the $\alpha+\alpha+t$ OCM
 to study the resonant structures of $^{11}$B.

%%%%%%%%%%%%%%%%%%%%%%%%%%%%%%%
\subsection{$\alpha+\alpha+t$ OCM with Gaussian expansion method}
%%%%%%%%%%%%%%%%%%%%%%%%%%%%%%%

The total wave function of $^{11}$B (the total angular momentum $J$ and total isospin $T=1/2$) 
 within the frame of the $\alpha+\alpha+t$ OCM is expanded in terms of the Gaussian basis,
\begin{eqnarray}
&&\Phi_J({^{11}{\rm B}})= \sum_c\sum_{\nu,\mu}A_c(\nu,\mu)\Phi_c^{(12,3)}(\nu,\mu)\nonumber \\
      && \hspace*{15mm} + \sum_c\sum_{\nu,\mu}B_c(\nu,\mu)\left[\Phi_c^{(23,1)}(\nu,\mu)+\Phi_c^{(31,2)}(\nu,\mu)\right],
\label{eq:total_wf}\\
&&\Phi_c^{(ij,k)}(\nu,\mu)=\left[\left[\varphi_{\ell}(\Vec{r}_{ij},\nu)\varphi_{\lambda}(\Vec{r}_k,\mu)\right]_L\chi_{\frac{1}{2}}(t)\right]_J,\\
&&\varphi_\ell(\Vec{r},\nu)=N_\ell(\nu)r^\ell \exp(-\nu r^2) Y_\ell (\hat{\Vec{r}}),
\end{eqnarray}
where we assign the cluster numbers as 1 and 2 for the two $\alpha$ clusters (spin 0)
 and 3 for $t$ cluster (spin 1/2).
$\Phi_J^{(12,3)}$ [$\Phi_J^{(23,1)+(31,2)}$] denotes the relative wave function of the $\alpha+\alpha+t$ system
 with the Jacobi-coordinate system shown in Fig.~\ref{fig:1}(a) [Figs.~\ref{fig:1}(b) and (c)].
It is noted that $\Phi_J^{(23,1)+(31,2)}$ is symmetric with respect to the particle-number exchange between 1 and 2.
$N_\ell$ is the normalization factor of the Gaussian basis $\varphi_\ell$, and $\Vec{r}_{ij}$ ($\Vec{r}_k$) denotes
 the relative coordinate between the {\it i}th and {\it j}th clusters (the {\it k}th cluster and
 the center-of-mass coordinate of the {\it i}th and {\it j}th clusters).   
The angular momentum channel is presented as $c=\left[(\ell,\lambda)_L\frac{1}{2}\right]_J$, where $\ell$ ($\lambda$) denotes 
 the relative orbital angular momentum with respect to $\Vec{r}_{ij}$ ($\Vec{r}_{k}$), and $L$ ($\frac{1}{2}$) is the total
 orbital angular momentum (the spin of $t$ particle).
The Gaussian parameter $\nu$ is taken to be of geometrical progression,
\begin{equation}
\nu_n=1/b_n^2,\hspace{1cm}b_n=b_{\rm min}a^{n-1},\hspace{1cm}n=1\sim n_{\max}.
\label{eq:para_Gaussian} 
\end{equation}
It is noted that this prescription is found to be very useful in optimizing
 the ranges with a small number of free parameters with high accuracy~\cite{kamimura88}.

The total Hamiltonian for the $\alpha+\alpha+t$ system is presented as
\begin{eqnarray}
\mathcal{H}&=&\sum_{i=1}^3 T_i-T_{cm} + V_{{2\alpha}}^{\rm (N)}(r_{12})+V_{{2\alpha}}^{\rm (C)}(r_{12})
           +\sum_{i=1}^2\left[V_{{\alpha}t}^{\rm (N)}(r_{i3})+V_{{\alpha}t}^{\rm (C)}(r_{i3})\right] \nonumber \\
              &+& V_{2\alpha t} + V_{\rm Pauli},
\label{eq:hamiltonian}
\end{eqnarray}
where $T_i$, $V_{2\alpha}^{\rm (N)}$ ($V_{{\alpha}t}^{\rm (N)}$) and $V_{2\alpha t}$
 stand for the kinetic energy operator
 for the {\it i}-th cluster, $\alpha$-$\alpha$ ($\alpha$-$t$) potential, and three-body potential, 
 respectively, and $V_{2\alpha}^{\rm (C)}$ ($V_{{\alpha}t}^{\rm (C)}$) is the Coulomb potential
 between 2$\alpha$ clusters ($\alpha$ and $t$).
The center-of-mass kinetic energy ($T_{\rm cm}$) is subtracted from the Hamiltonian.
The Pauli-blocking operator $V_{\rm Pauli}$~\cite{kukulin84} is expressed as
\begin{eqnarray}
&&V_{\rm Pauli}=\lim_{\lambda \rightarrow \infty} \lambda \hat{O}_{\rm Pauli},\label{eq_A:Pauli}\\ 
&&\hat{O}_{\rm Pauli}
             = \sum_{2n+\ell<4,\ell=even}
               \left|u_{n\ell}(\Vec{r}_{12}\rangle\langle u_{n\ell}(\Vec{r}_{12})\right|
             + \sum_{2n+\ell<3}\sum_{i=1}^2
               \left|u_{n\ell}(\Vec{r}_{i3}\rangle\langle u_{n\ell}(\Vec{r}_{i3})\right|,
               \label{Pauli_operator}
\end{eqnarray}
which removes the Pauli forbidden states between the two $\alpha$-particles
 in $0S$, $0D$ and $1S$ states as well as those between $\alpha$ and $t$ particles
 in $0S$, $0P$, $0D$ and $1S$ states.
Consequently, the ground state with the shell-model-like configuration
 can be described properly in the present model.  
The effective $\alpha$-$\alpha$ potential and Coulomb potential, $V_{2\alpha}^{\rm (N)}$
 and $V_{2\alpha}^{\rm (C)}$, 
 are constructed with the folding procedure, where we fold the modified Hasegawa-Nagata 
 effective $NN$ interaction (MHN)~\cite{mhn} and the $pp$ Coulomb potential with the $\alpha$-particle density.
They reproduce the observed $\alpha\alpha$ scattering phase shifts ($S$-, $D$- and $G$-waves)
 and the energies of the $^{8}$Be ground state and
 of the Hoyle state ($0_2^+$ in $^{12}$C)~\cite{yamada05,funaki08}.  
Concerning the effective $\alpha$-${\it t}$ potential and Coulomb potentials, we use the folding
 potentials which reproduces well the low-lying energy spectra of $^7$Li [$3/2^-_1$ (g.s.), $1/2^-_1$,
 $7/2^-_1$, and $5/2^-_1$] (including the low-energy $\alpha-t$ scattering phase shifts)
 obtained initially by Nishioka et al.~\cite{nishioka79,furutani80}. 
The potentials are applied to the structure study of $^7_\Lambda$Li with the $\alpha$+$t$+$\Lambda$
 model and are known to reproduce well the low-lying structure of the hypernucleus~\cite{hiyama97,hiyama09}.
It is reminded that the $\alpha-t$ potential has a strong parity-dependence:~ The odd-parity potentials are
 attractive to reproduce the bound states ($3/2^{-}$, $1/2^{-}$) with respect to the $\alpha+t$ threshold, 
 while the even-parity potentials are weakly attractive and thus no bound/resonant states with positive 
 parity have not been observed in low-energy excitation region.
 
The three-body potential $V_{2\alpha{\it t}}$ is introduced phenomenologically so as
 to reproduce the energies of the ground state ($3/2^-_1$) and the first excited 
 positive parity state ($1/2^+_1$) of $^{11}$B with respect to the $\alpha+\alpha+t$ threshold.
The origin of $V_{2\alpha{\it t}}$ is considered to derive from the state dependence of the effective 
 nucleon-nucleon interaction and an additional Pauli repulsion arising from exchanging nucleons
 among the three clusters ($\alpha+\alpha+t$). 
It should be short-range, and hence only act in compact configurations.
In the present paper, we take the following phenomenological three-body potential,
\begin{eqnarray}
V_{2\alpha{\it t}} =  \sum_{Q=7,8} V_0(Q) \sum_{(\lambda\mu)} \sum_{L^{\pi}}
  {|\Phi^{\rm SU(3)}_{(\lambda\mu)Q}({L^{\pi}})\rangle} {\langle\Phi^{\rm SU(3)}_{(\lambda\mu)Q}(L^{\pi})|},
\end{eqnarray}
where $\Phi^{\rm SU(3)}_{(\lambda\mu)Q}({L^{\pi}})$ with the total orbital angular momentum $L$ represents
 the SU(3)$[443](\lambda\mu)$ wave function with the total oscillator quanta $Q$ ($Q\geq 7$).
It is noted that the present $\alpha+\alpha+t$ model space can be classified into 
 the SU(3) bases with the irreducible representation, $(\lambda\mu)Q$, with partition $[f]=[443]$,
 and the total wave function of $^{11}$B with positive (negative) parity in Eq.~(\ref{eq:total_wf}) 
 can be expanded in terms of the $\Phi^{\rm SU(3)}_{(\lambda\mu)Q}({L^{\pi}})$ bases
 with even (odd) $Q$ values~\cite{nishioka79,furutani80}.
The $(\lambda\mu)=(13)$ basis with $Q=7$ is unique Pauli allowed state for negative-parity
 states of $^{11}$B with $Q=7$, which is equivalent to the shell-model configuration of $(0s)^4(0p)^7$.
Thus this SU(3) basis becomes the main component in the ground state of $^{11}$B.
On the other hand, there exist only three Pauli-allowed states, $(\lambda\mu)=(42)$, $(23)$ and $(04)$, 
 for the positive-parity states with $Q=8$.
Therefore they correspond to the main components in the first excited states with even parity in the present model.
For simplicity, the strengths of the three-body potential, $V_0(Q=7)$ and $V_0(Q=8)$, are fixed so as to reproduce
 the experimental energies of the ground state of $^{11}$B and
 the first $1/2^+$ state ($E_x=6.79$ MeV), respectively, with respect to the $\alpha+\alpha+t$ threshold.
The expectation value of this three-body potential does not exceed 10~\% of that of the corresponding two-body term, 
 even for the ground state with the most compact structure, i.e.~being the most sensitive to the potential.  

The equation of motion of $^{11}$B with the $\alpha+\alpha+t$ OCM is obtained by the variational principle,
\begin{eqnarray}
\delta\left[\langle\Phi_J({^{11}{\rm B}})\mid \mathcal{H}-E \mid \Phi_J({^{11}{\rm B}})\rangle\right]=0,
\label{eq:variational_principle}
\end{eqnarray}
where $E$ denotes the eigenenergy of $^{11}$B measured from the $\alpha+\alpha+t$ threshold.
The energy $E$ and expansion coefficients $A_c$ and $B_c$ in the total wave function shown
 in Eq.~(\ref{eq:total_wf}) are determined by solving a secular equation derived from 
 Eq.~(\ref{eq:variational_principle}). 

It is instructive to study single-$\alpha$-particle (single-$t$-particle) orbits and corresponding occupation
 probabilities in $^{11}$B. 
We define the single-cluster density matrices for $\alpha$ and $t$ clusters, respectively, as
\begin{eqnarray}
&&\rho^{(\alpha)}(\Vec{r},\Vec{r}')
 = \langle \Phi_J({^{11}{\rm B}}) |~\frac{1}{2}\sum_{i=1}^2 {|\delta(\Vec{r}^{(G)}_i-\Vec{r}')\rangle}{\langle\delta(\Vec{r}^{(G)}_i-\Vec{r})|}~|\Phi_J({^{11}{\rm B}})\rangle,
\label{eq:single_alpha_density}\\
&&\rho^{(t)}(\Vec{r},\Vec{r}')
 = \langle \Phi_J({^{11}{\rm B}}) |~{|\delta(\Vec{r}^{(G)}_3-\Vec{r}')\rangle}{\langle\delta(\Vec{r}^{(G)}_3-\Vec{r})|}~|\Phi_J({^{11}{\rm B}})\rangle,
\label{eq:single_triton_density}
\end{eqnarray}
where $\Vec{r}^{(G)}_i$ $(i=1,2)$ [$\Vec{r}^{(G)}_3$] represents the coordinate vector of
the $i$th $\alpha$ (triton) cluster with respect to the center-of-mass coordinate of the $\alpha+\alpha+t$ system. 
The single-$\alpha$-particle (single-$t$-particle) orbits and corresponding occupation probabilities are obtained 
 by solving the eigenvalue equation of the single-cluster density matrices,
\begin{eqnarray}
&&\int d\Vec{r} \rho^{(\alpha)}(\Vec{r},\Vec{r}') f^{(\alpha)}_\mu(\Vec{r}') = \mu^{(\alpha)} f^{(\alpha)}_\mu(\Vec{r}),
\label{eq:density_alpha}\\
&&\int d\Vec{r} \rho^{(t)}(\Vec{r},\Vec{r}') f^{(t)}_\mu(\Vec{r}') = \mu^{(t)} f^{(t)}_\mu(\Vec{r}),
\label{eq:density_triton}
\end{eqnarray}
where the eigenvalue $\mu^{(\alpha)}$ ($\mu^{(t)}$) denotes the occupation probability for the corresponding
 single-cluster orbit $f^{(\alpha)}_\mu$ ($f^{(t)}_\mu$) with the argument of the intrinsic coordinate
 of an arbitrary $\alpha$ (triton) cluster in a nucleus measured from the center-of-mass
 coordinate of $^{11}$B.
The spectrum of the occupation probabilities provide important information on the occupancies
 of the single-$\alpha$- ($t$-) particle orbit in $^{11}$B.
If the two $\alpha$ particles (one $t$ particle) occupy only an single orbit,
 the occupation probability for this orbit becomes 100~\%.

The nuclear root-mean-square (rms) radius of $^{11}$B in the present OCM is given as
\begin{eqnarray}
R_{N} = \left[ \frac{1}{11}\left( 2{\langle r^2 \rangle}_{\alpha} + {\langle r^2 \rangle}_{t} + 2{R_{\alpha-\alpha}}^2 + \frac{24}{11}{R_{{^8{\rm Be}}-t}}^2 \right) \right]^{1/2},
\label{eq:rms}
\end{eqnarray}
where $R_{\alpha-\alpha}$ ($R_{{^8{\rm Be}}-t}$) presents the rms distance
 between $\alpha$ and $\alpha$ ($^8{\rm Be}$ and $t$) in $^{11}$B.
In Eq.~(\ref{eq:rms}) we take into account the finite size effect of $\alpha$ and $t$ clusters, where the experimental rms radii
 for the $\alpha$ and $t$ particles are used in $\sqrt{\langle r^2 \rangle}_{\alpha}$ and
 $\sqrt{\langle r^2 \rangle}_{t}$, respectively.

The reduced width amplitude or overlap amplitude is useful to see the degree of clustering in nucleus.
In the present paper, we study the reduced width amplitudes for the $^7$Li+$\alpha$ and $^8$Be+t channels,
 respectively, defined as
\begin{eqnarray}
&&{\cal Y}_{J_7\ell_{74}J}^{{^7}{\rm Li}-\alpha}({r}_{74})
 = r_{74} \times \left\langle \left[ \frac{\delta({r^{\prime}_{74}}-r_{74})}{{r^{\prime}_{74}}^2} \phi_{J_7}{({^7}{\rm Li})} Y_{\ell_{74}}(\hat{\Vec{r}}^{\prime}_{74})\right]_J | \Phi_J({^{11}{\rm B}}) \right\rangle,
\label{eq:RWA_alpha}\\
&&{\cal Y}_{J_8(\ell_{83}\frac{1}{2})j_{83}J}^{{^8}{\rm Be}-t}({r}_{83})
 = r_{83} \times \left\langle \left[ \frac{\delta({r^{\prime}_{83}}-r_{83})}{{r^{\prime}_{83}}^2} \phi_{J_8}({^8}{\rm Be}) [Y_{\ell_{83}}(\hat{\Vec{r}}^{\prime}_{83})\chi_{\frac{1}{2}}(t)]_{j_{83}}\right]_J | \Phi_J({^{11}{\rm B}}) \right\rangle,
\label{eq:RWA_triton}
\end{eqnarray}
where $r_{74}$ ($r_{83}$) denotes the radial part of the relative coordinate between $^{7}$Li and $\alpha$ ($^8$Be and $t$).
The wave function $\phi_{J_7}{({^7}{\rm Li})}$ ($\phi_{J_8}({^8}{\rm Be})$) of $^{7}$Li ($^{8}$Be) 
 with the total angular momentum $J_7$ ($J_8$) is obtained with the $\alpha+t$ OCM ($\alpha+\alpha$ OCM). 
The spectroscopic factor $S^2$ is defined as 
\begin{eqnarray}
S^{2}=\int_{0}^{\infty} dr [{\cal Y}(r)]^2,
\end{eqnarray}
where ${\cal Y}$ denotes the reduced width amplitude.

%%%%%%%%%%%%%%%%%%%%%%%%%%%%%%%%%%%%%%%%%%%%%%%%%%%%%%%%%%%%%%%%%%%%%%%%%%%%%%%%
\subsection{Complex-scaling method with $\alpha+\alpha+t$ OCM}\label{sec:CSM}
%%%%%%%%%%%%%%%%%%%%%%%%%%%%%%%%%%%%%%%%%%%%%%%%%%%%%%%%%%%%%%%%%%%%%%%%%%%%%%%%
 
In the present study, we take the complex-scaling method (CSM) for studying the resonant structures of $^{11}$B.
The CSM is a powerful tool to investigate the resonant parameters 
 (energies and widths) of the resonant states as well as the binding energies of the bound states~\cite{aguilar71}.
Application of CSM is easy in the framework of the $\alpha+\alpha+t$ OCM.

In CSM, we transform the relative coordinates of the $\alpha+\alpha+t$ model shown in Fig.~\ref{fig:1},
 $\{\Vec{r}_{ij},\Vec{r}_{k}\}$ with $(ij,k)$=(12,3), (23,1) and (32,1), by the operator $U^{\theta}$ as
\begin{eqnarray} 
U^{\theta}\Vec{r}_{ij} = \Vec{r}_{ij}e^{i\theta}~~{\rm and}~~U^{\theta}\Vec{r}_{k} = \Vec{r}_{k}e^{i\theta},
\end{eqnarray} 
where $\theta$ is a scaling angle. 
Then, the Hamiltonian in Eq.~(\ref{eq:hamiltonian}) is transformed into the complex-scaled Hamiltonian,
\begin{eqnarray} 
H^{\theta}=U^{\theta}H(U^{\theta})^{-1}. 
\label{eq:complex_scaled_Hamiltonian}
\end{eqnarray} 
The corresponding complex-scaled Schr\"odinger equation
 is expressed as
\begin{eqnarray} 
&&H^{\theta} \Phi^{\theta}_{J} = E \Phi^{\theta}_{J},  \label{eq:CSM}\\ 
&&\Phi^{\theta}_{J}=e^{(3/2)i\theta\times 2} \Phi_{J}(\{\Vec{r}_{ij}e^{i\theta},\Vec{r}_{k}e^{i\theta}\})
\label{eq:CSM_wf}
\end{eqnarray} 
 where $\Phi_{J}$ is given in Eq.~(\ref{eq:total_wf}). 
The eigenstates $\Phi^{\theta}_{J}$ are obtained by solving the eigenvalue problem of $H^{\theta}$
 in Eq.~(\ref{eq:CSM}). 
In CSM, all the energy eigenvalues $E$ of bound and unbound states are obtained
 on a complex energy plane, according to ABC theorem~\cite{aguilar71}. 
In this theorem, it is proven that the boundary condition of Gamow resonances is transformed
 to the damping behavior at the asymptotic region. 
Thanks to this condition, the same theoretical method as that used for the bound states
 can be employed to obtain the many-body resonances. 
For $^{11}$B, the continuum states of $^7$Li+$\alpha$, $^8$Be+$t$ and $\alpha+\alpha+t$ channels
 are obtained on the branch cuts rotated with the $2\theta$ dependence~\cite{aguilar71}. 
On the contrary, bound states and resonances are discrete and obtainable independently of $\theta$.
Thus, these discrete states are located separately from the many-body continuum spectra
 on the complex energy plane. 
One can identify the resonance poles of complex eigenvalues: $E = E_{r} - i\Gamma/2$, 
 where $E_{r}$ and $\Gamma$ are the resonance energy measured from the $\alpha+\alpha+t$ threshold
 and the decay width, respectively.
 
In order to solve the complex-scaled Schr\"odinger equation Eq.~(\ref{eq:CSM}), we expand the wave function
 Eq.~($\ref{eq:CSM_wf}$) in terms of the Gaussian basis functions shown in Eq.~(\ref{eq:total_wf}),
\begin{eqnarray}
&&\Phi^{\theta}_{J} = \sum_c\sum_{\nu,\mu}F^{\theta}_c(\nu,\mu)\Phi_c^{(12,3)}(\nu,\mu)\nonumber \\
      && \hspace*{15mm} + \sum_c\sum_{\nu,\mu}G^{\theta}_c(\nu,\mu)\left[\Phi_c^{(23,1)}(\nu,\mu)+\Phi_c^{(31,2)}(\nu,\mu)\right].
\label{eq:wf_CSM_Gaussian}
\end{eqnarray} 
The expansion coefficients $F^{\theta}_c(\nu,\mu)$ and $G^{\theta}_c(\nu,\mu)$ together with the discrete energy spectrum $E$
 are obtained by solving an eigenvalue problem derived from
 Eqs.~(\ref{eq:CSM}) and (\ref{eq:wf_CSM_Gaussian})~\cite{kuruppa88,kuruppa90,aoyama06}.
This complex scaling method with the Gaussian basis functions is known to give high precision needed
 to maintain the numerical stability of the complex eigenvalue problem.

%%%%%%%%%%%%%%%%%%%%%%%%%%%%%%%%%%%%%%%%%%%%%%%%%%%%%%%%%%%%%%%%%%
\section{Results and discussion}\label{sec:results_discussions}
%%%%%%%%%%%%%%%%%%%%%%%%%%%%%%%%%%%%%%%%%%%%%%%%%%%%%%%%%%%%%%%%%%

\begin{figure}[tbh]
\begin{center}
\epsfig{file=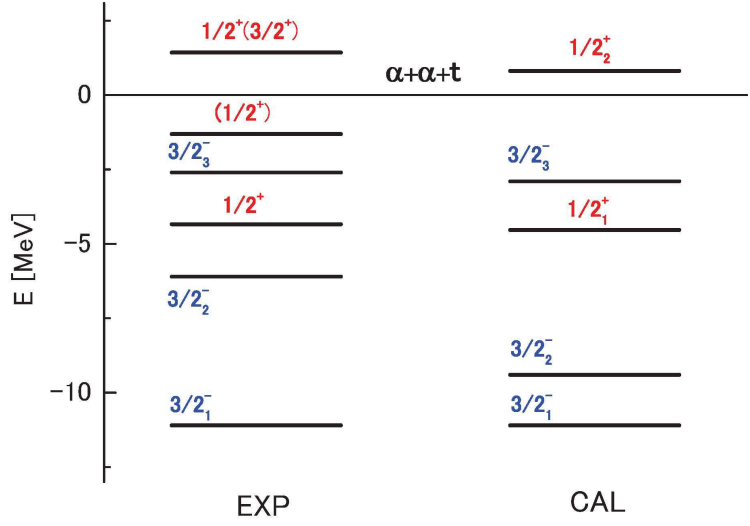,width=0.6\hsize}
\caption{
Calculated energy levels of $3/2^{-}$ and $1/2^+$ states in $^{11}$B 
 with respect to the $\alpha+\alpha+t$ threshold together
 with the experimental data~\cite{ajzenberg86}.
}
\label{fig:2}
\end{center}
\end{figure}

The energy levels of $3/2^{-}$ and $1/2^+$ states in $^{11}$B with respect to the $\alpha+\alpha+t$ threshold
 are shown in Fig.~\ref{fig:2}, which are obtained with the $\alpha+\alpha+t$ OCM using GEM.
In the present calculation, we found that three $3/2^-$ states and two $1/2^+$ states come out 
 as either bound states against particle decays or resonant states.  

%%%%%%%%%%%% 
%  Fig. 2  %
%%%%%%%%%%%%

%%%%%%%%%%%%%%%%%%%%%%%%%%%%%%%%%
\subsection{$3/2^{-}$ states}\label{sec:3/2-}
%%%%%%%%%%%%%%%%%%%%%%%%%%%%%%%%%

First we discuss the structures of the three $3/2^{-}$ states.
The $3/2^{-}_{1}$ state at $E=-11.1$ MeV measured from the $\alpha+\alpha+t$ threshold
 corresponds to the ground state of $^{11}$B.
The calculated nuclear radius is $R_{N}$=2.22~fm (see Table~\ref{tab:1}),
 the value of which is in correspondence with the experimental data $2.43\pm0.11$~fm~\cite{ajzenberg86}.
According to the analysis of the wave function, the main component of this state 
 is SU(3)$[f](\lambda,\mu)_{L}=[443](1,3)_{1}$ with $Q=7$ harmonic oscillator quanta (95~\%) 
 and its dominant angular momentum channel is $(L,S)_{J}=(1,\frac{1}{2})_{\frac{3}{2}}$. 
Thus, this state has a compact shell-model-like structure, in which the two $\alpha$ and one $t$ clusters
 are heavily overlapping and melting away each other (due to antisymmetrization among nucleons)
 to have a shell-model configuration.
On the other hand, we found that the $3/2^{-}_{2}$ state at $E=-9.4$ MeV has also a compact
 shell-model-like structure with the dominant component of SU(3)$(\lambda,\mu)_{L}=(1,3)_{2}$ with $Q=7$
 and $(L,S)_{J}=(2,\frac{1}{2})_{\frac{3}{2}}$.
The nuclear radius of this state is $R_{N}$=2.23~fm, the value of which is similar to that of the ground state.
The energy difference between the $3/2^{-}_{1}$ and $3/2^{-}_{2}$ states is about $2$ MeV,
 which is smaller than the experimental data ($\sim 5$~MeV).    
This feature is recovered if we take into account the SU(3)$[4421]$ shell model configuration
 in the present OCM model space
 (or adopting an extended $\alpha+\alpha+d+n$ model space in the cluster-model terminology),
 as pointed out by Nishioka et al~\cite{nishioka79,furutani80}.

%%%%%%%%%%%%%%%%%%%%%%%%%%%%%%%%%%%%%%%%%%%%%%%%%%%
%  TABLE 1  %
%%%%%%%%%%%%%%%%%%%%%%%%%%%%%%%%%%%%%%%%%%%%%%%%%%%
\begin{table}
\caption{
Calculated nuclear radii ($R_{N}$), 
 $\alpha-\alpha$ rms distances ($R_{\alpha-\alpha}$) and ${^8}{\rm Be}-t$
 ones ($R_{{{^8}{\rm Be}}-t}$) for the $3/2^{-}$ states 
 of $^{11}$B together with their $\alpha-t$ and ${{^7}{\rm Li}}-\alpha$ rms distances
 ($R_{\alpha-t}$ and $R_{{{^7}{\rm Li}}-\alpha}$, respectively).
They are given in units of fm. 
}
\label{tab:1}
\begin{center}
\begin{tabular}{cccccc}
\hline
\hline
\hspace{10mm}$J^\pi$\hspace{8mm} 
     & \hspace{3mm}{$R_{N}$}\hspace{3mm} 
     & \hspace{3mm}{$R_{\alpha-\alpha}$}\hspace{3mm}
     & \hspace{3mm}{$R_{{{^8}{\rm Be}}-t}$}\hspace{3mm} 
     & \hspace{3mm}{$R_{\alpha-t}$}\hspace{3mm}
     & \hspace{3mm}{$R_{{{^7}{\rm Li}}-\alpha}$}\hspace{3mm}\\
\hline
 $3/2^{-}_{1}$ & $2.22$  & $2.45$ & ~~$2.14$ & $2.47$ & ~~$2.13$ \\ 
 $3/2^{-}_{2}$ & $2.23$  & $2.49$ & ~~$2.17$ & $2.51$ & ~~$2.17$ \\[1mm]
 $3/2^{-}_{3}$ & $3.00$  & $4.47$ & ~~$3.49$ & $4.15$ & ~~$3.82$ \\ 
\hline
\hline
\end{tabular}
\end{center}
\end{table}
%%%%%%%%%%%%%%%%%%%%%%%%%%%%%%%%%%%%%%%%%%%%%%%%%%%

In addition to the two $3/2^{-}$ states discussed above, the $3/2^{-}_{3}$ state appears at 
 $E_x=8.2$~MeV ($E=-2.9$~MeV with respect to the $\alpha+\alpha+t$ threshold).
The nuclear radius of $3/2^{-}_3$ is $R_{N}$=3.00~fm.
This value is by about $30$~\% larger than that of the ground state of $^{11}$B
 and is consistent with the AMD calculation ($R_{N}$=3.0~fm)~\cite{enyo07}.
The rms distance between $\alpha$ and $\alpha$ (between $^{8}$Be($2\alpha$) and $t$) is 
 $R_{\alpha-\alpha}=4.47$~fm ($R_{^8{\rm Be}-t}=3.49$~fm).
It is reminded that the value of $R_{\alpha-\alpha}\sim 4$~fm in the ground state of $^8$Be
 with the $\alpha+\alpha$ OCM calculation.
Thus, the intra-cluster distance between the two $\alpha$'s in $3/2^{-}_{3}$ is 
 a little bit larger that that in the ground state of $^{8}$Be.
In addition, the intra-cluster distance between the $\alpha$ and $t$ clusters
 in $3/2^{-}_3$ is 4.15~fm (see Table~\ref{tab:1}), which is also larger than
 that ($3.5$~fm) in the ground state of $^{7}$Li($\alpha+t$). 

\begin{figure}[tbh]
\begin{tabular}{cc}
\begin{minipage}{0.45\hsize}
\begin{center}
\epsfig{file=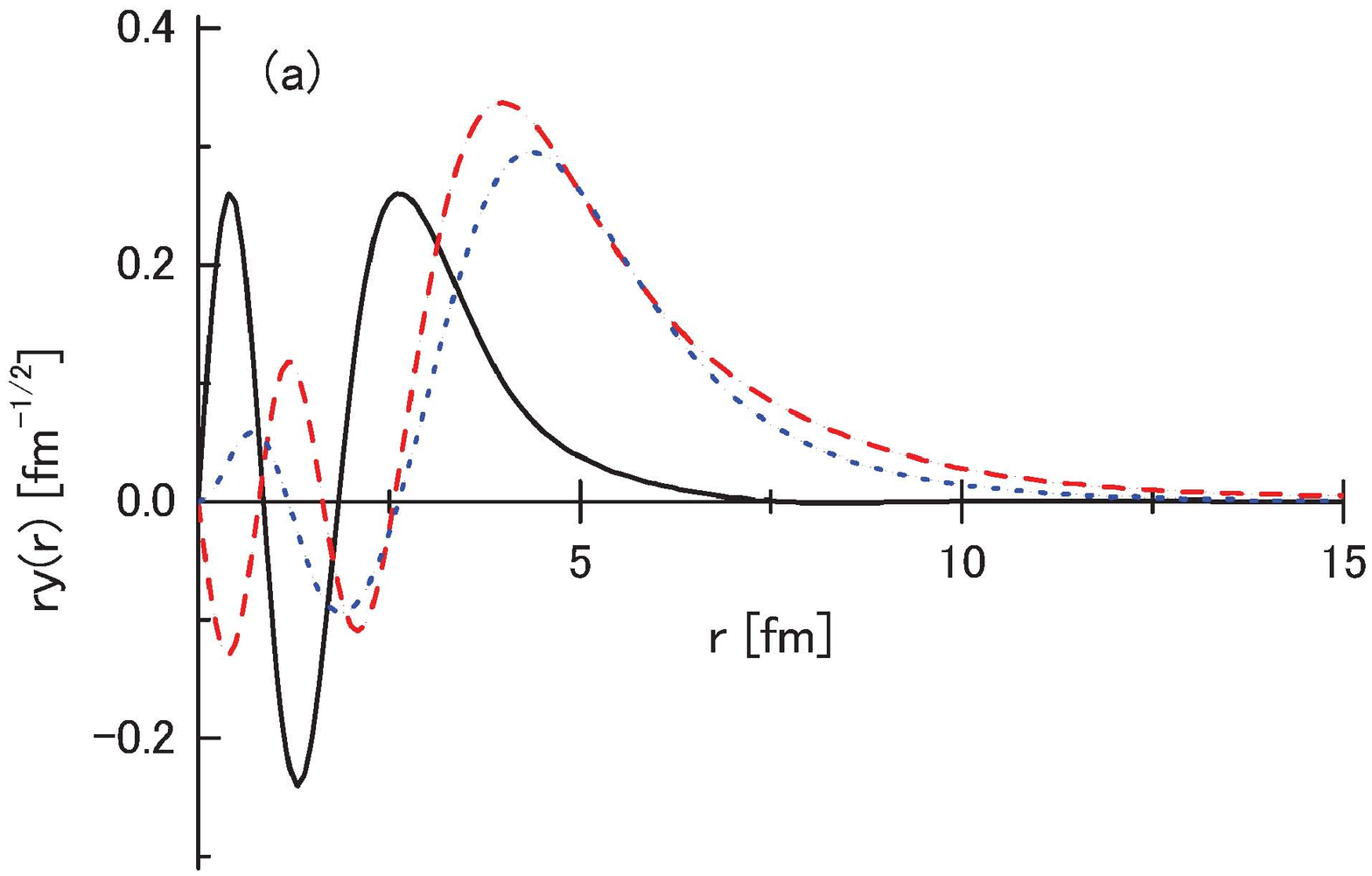,width=\hsize}
\end{center}
\end{minipage}
\hspace*{2mm}
\begin{minipage}{0.45\hsize}
\begin{center}
\epsfig{file=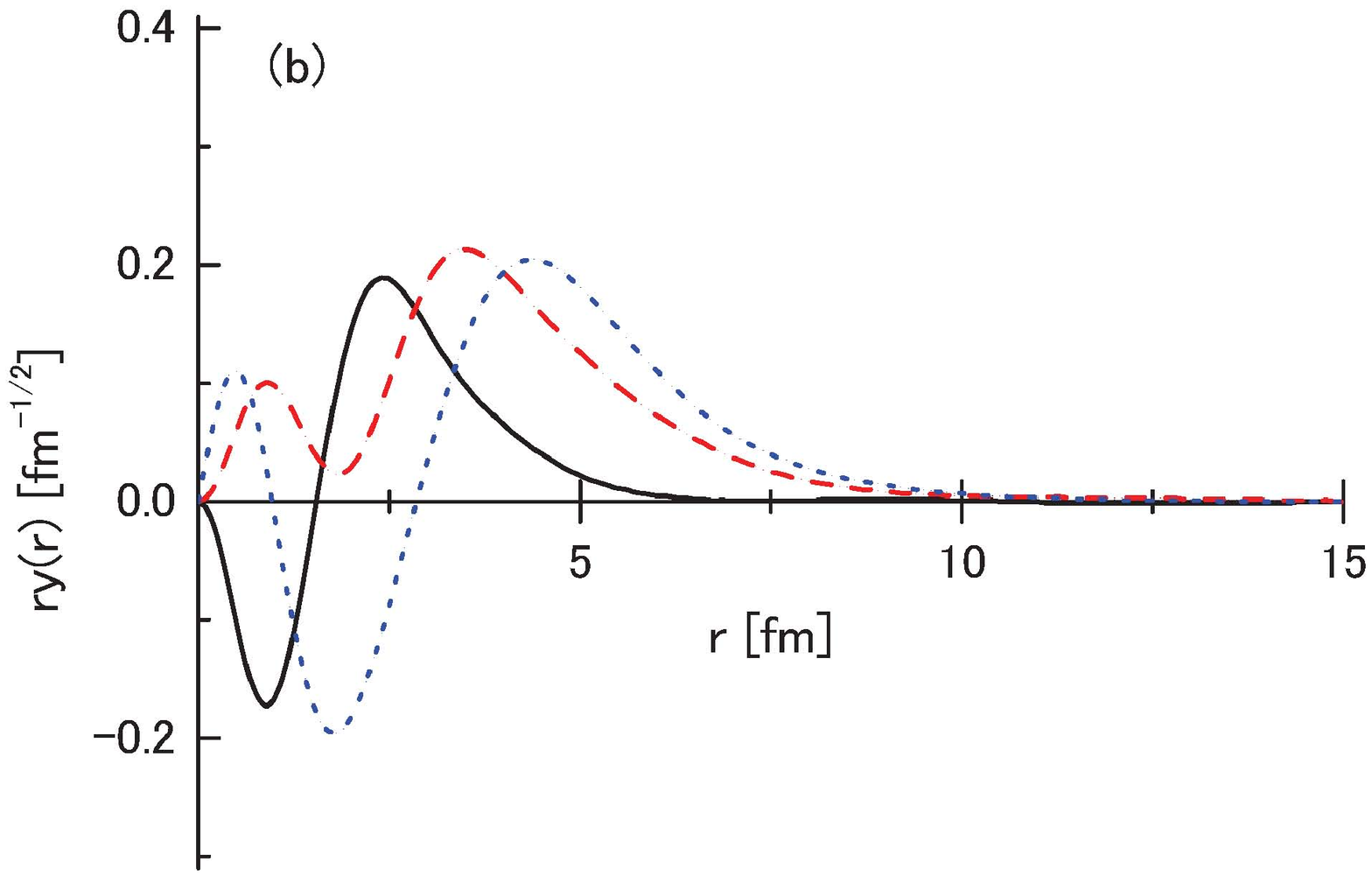,width=\hsize}
\end{center}
\end{minipage}
\end{tabular}
\begin{center}
\caption{
Reduced width amplitudes or overlap amplitudes of
 the (a)~$^{7}$Li($3/2^-$;~g.s)+$\alpha$ and (b)~$^{8}$Be($0^+$;~g.s)+$t$
 channels in the $3/2^{-}_1$ (real line) 
 and $3/2^{-}_3$ (dashed) states of $^{11}$B.
The relative angular momentum of the amplitudes are $\ell_{74}=0$ for (a)
 and $j_{83}=3/2^{-}$ for (b).
The dotted lines in (a) and (b) present, respectively, the reduced width amplitudes of 
 $^{7}$Li($3/2^-$;~g.s)+$\alpha$ ($\ell=1$) and $^{8}$Be($0^+$:~g.s)+$t$ ($\ell=0$) in the $1/2^+_1$ state.
}
\label{fig:3}
\end{center}
\end{figure}

In order to study the structure of $3/2^{-}_3$, it is interesting to compare the reduced
 width amplitudes or overlap amplitudes of
 the $^{7}$Li(g.s)+$\alpha$ channel [$^{8}$Be(g.s)+$t$] defined in Eq.~(\ref{eq:RWA_alpha})
 [(\ref{eq:RWA_triton})] for the $3/2^{-}_3$ and $3/2^{-}_1$ states. 
The results are shown in Fig.~\ref{fig:3}. 
One sees that 1)~the reduced width amplitudes of the $^{7}$Li+$\alpha$ and $^{8}$Be+$t$ channels
 in $3/2^{-}_3$ are enhanced and significantly larger than those in the ground state, in particular, 
 in the outermost-peak region, and 2)~the number of the nodes of their amplitudes in the former state
 increase by one larger than those in the latter.
It is noted that the radial behaviors of the reduced width amplitudes in the ground state
 of $^{11}$B are determined by the dominant nature of SU(3)$(\lambda,\mu)=(1,3)$
 in its state.    
As for the $S^2$ factors in $3/2^{-}_{3}$, 
 the $^7$Li(g.s)+$\alpha$ channel with the relative orbital angular momentum $\ell_{74}=0$
 has the largest value of $0.278$ among all of the $^7$Li+$\alpha$ channels,
 and the secondarily (thirdly) largest channel
 is the $^7$Li(g.s)+$\alpha$ ($^7$Li($1/2^-$)+$\alpha$) channel ($\ell_{74}=2$)
 with $S^2=0.126$ (0.125),
 while the $^8$Be(g.s)+$t$ channel with the relative angular momentum $j_{83}=3/2$
 has the largest value of $0.102$.
In the $3/2^{-}_{1}$ state, the corresponding $S^2$ factors are $0.144$, $0.062$, 
 $0.052$ and $0.062$, respectively for $^7$Li(g.s)+$\alpha$ ($\ell_{74}=0$), 
 $^7$Li(g.s)+$\alpha$ ($\ell_{74}=2$),  $^7$Li($1/2^{-}$)+$\alpha$ ($\ell_{74}=2$),
 and $^8$Be(g.s)+$t$ ($j_{83}=3/2$). 
Consequently, the $3/2^{-}_{3}$ state has an $\alpha+\alpha+t$
 cluster structure, although the cluster structure of the $\alpha$+$t$ part
 is significantly distorted from that of the ground state of $^7$Li.

%%%%%%%%%%%%%%%%%%%%%%%%%%%%%%%%%%%%%%%%%%%%%%%%%%%
%  Fig. 3  %
%%%%%%%%%%%%%%%%%%%%%%%%%%%%%%%%%%%%%%%%%%%%%%%%%%%
%%%%%%%%%%%%%%%%%%%%%%%%%%%%%%%%%%%%%%%%%%%%%%%

Recently, Kawabata et al.~measured the isoscalar monopole transition rate $B$(E0:IS) for
 $3/2^{-}_{3}$ with the $^{11}$B$(d,d')$ reaction~\cite{kawabata07}.
The experimental value is as large as $96\pm 16$~fm$^4$, comparable to that of the Hoyle state
 in $^{12}$C ($B$(E0:IS)$=120\pm 9$~fm$^4$).
The calculated result in the present model is $B$(E0:IS)=$92$~fm$^4$,
 in good agreement with the experimental data. 
On the other hand, the AMD calculation also reproduces the experimental
 data ($B^{\rm AMD}$(E0:IS)=$94$~fm$^4$).
They analyzed the density distributions of the ground state and excited states
 in $^{11}$B and $^{12}$C, and concluded that the $3/2^{-}_{3}$ state in $^{11}$B
 is a dilute cluster state, and its features are similar to those of the Hoyle
 state $^{12}$C$(0^+_2)$ which is likely to be a gas state of $3\alpha$ clusters.
However, it is not self-evident in the AMD calculation whether
 the $3/2^{-}_{3}$ state possesses an $\alpha$ condensate nature like the Hoyle state having
 about $70$\% probability of the $\alpha$ particle occupied in a single $S$ orbit.
In order to see the single-cluster properties in $^{11}$B, we study the single-cluster
 orbits and their occupation probabilities in the $3/2^{-}_{3}$ state by solving
 the eigenvalue equation of the single-cluster density matrices in Eqs.~(\ref{eq:density_alpha})
 and (\ref{eq:density_triton}), and compare them with those in the Hoyle state.
 
%%%%%%%%%%%% 
%  Fig. 4  %
%%%%%%%%%%%%
\begin{figure}[tbh]
\begin{tabular}{cc}
\begin{minipage}{0.45\hsize}
\begin{center}
\epsfig{file=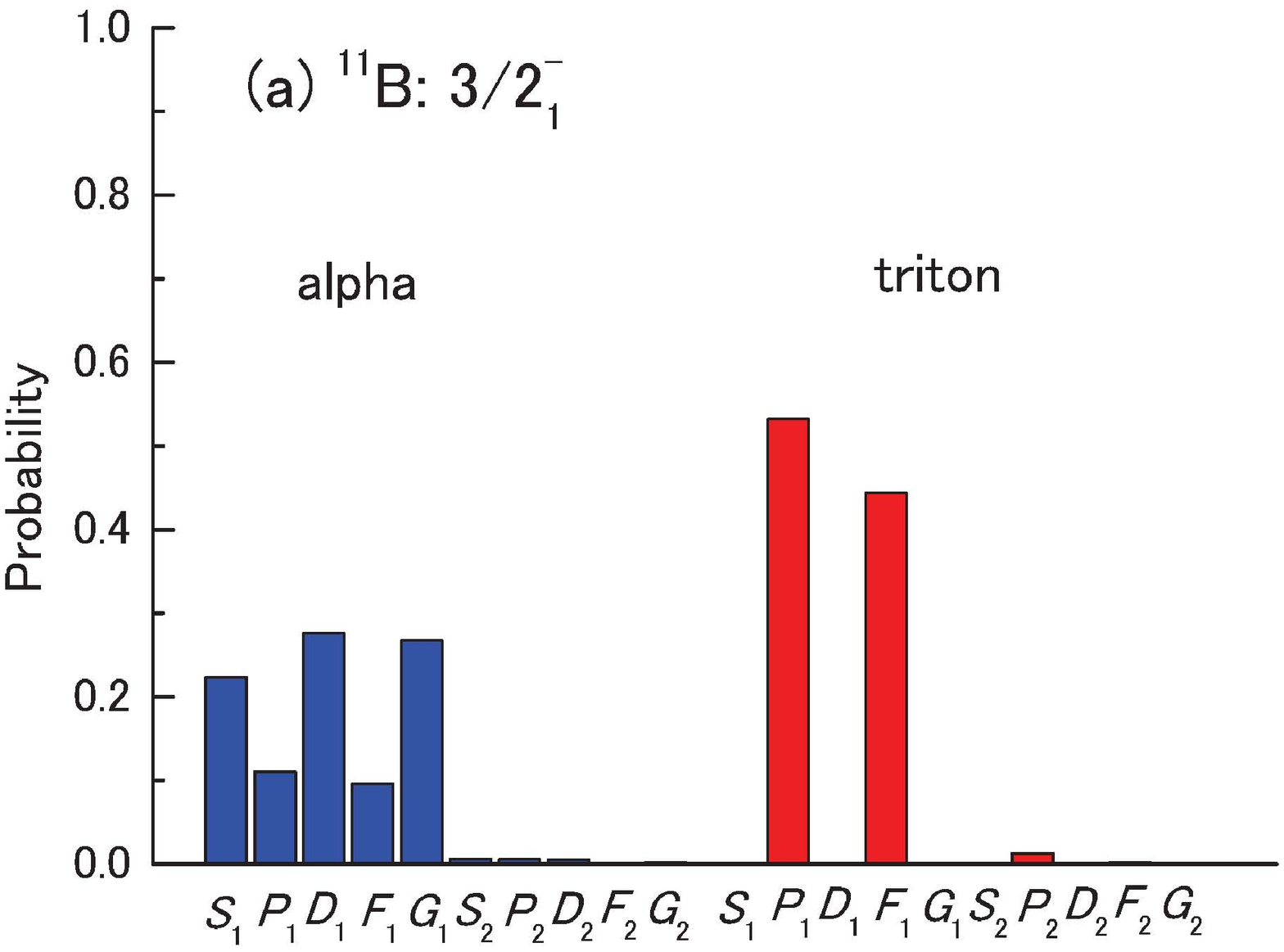,width=\hsize}
\end{center}
\end{minipage}
\hspace*{2mm}
\begin{minipage}{0.45\hsize}
\begin{center}
\epsfig{file=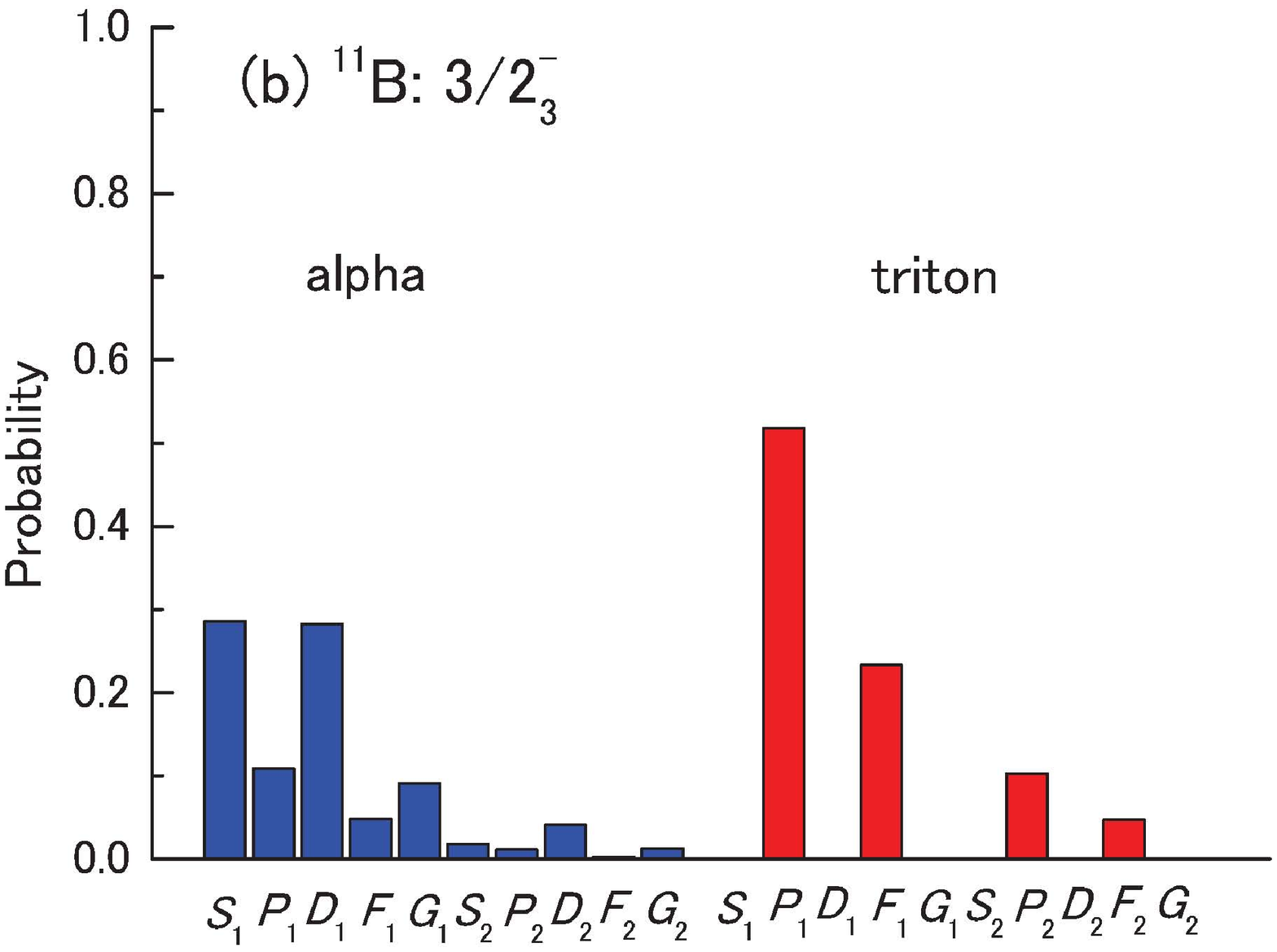,width=\hsize}
\end{center}
\end{minipage}
\\\\
\begin{minipage}{0.45\hsize}
\begin{center}
\epsfig{file=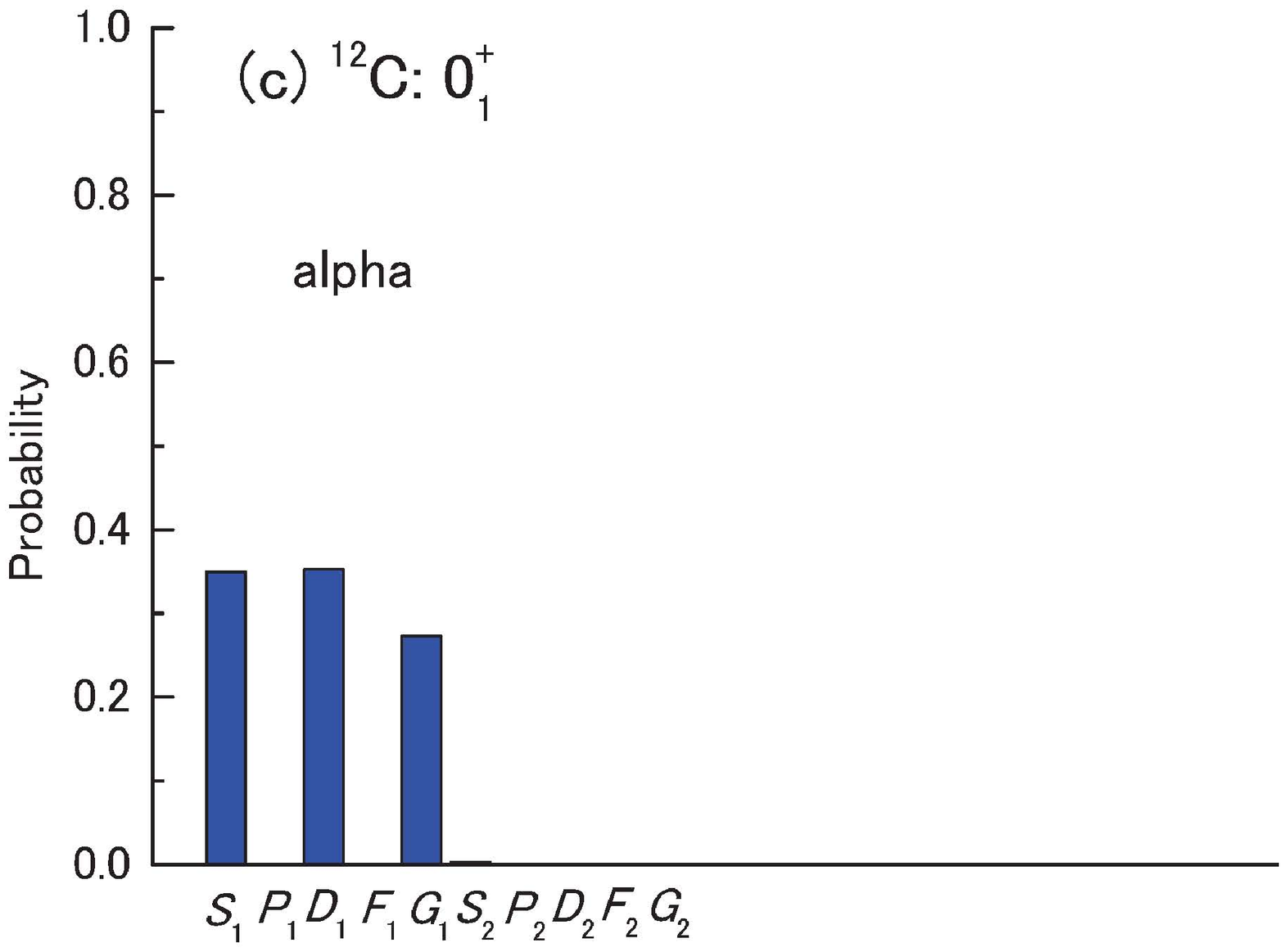,width=\hsize}
\end{center}
\end{minipage}
\hspace*{2mm}
\begin{minipage}{0.45\hsize}
\begin{center}
\epsfig{file=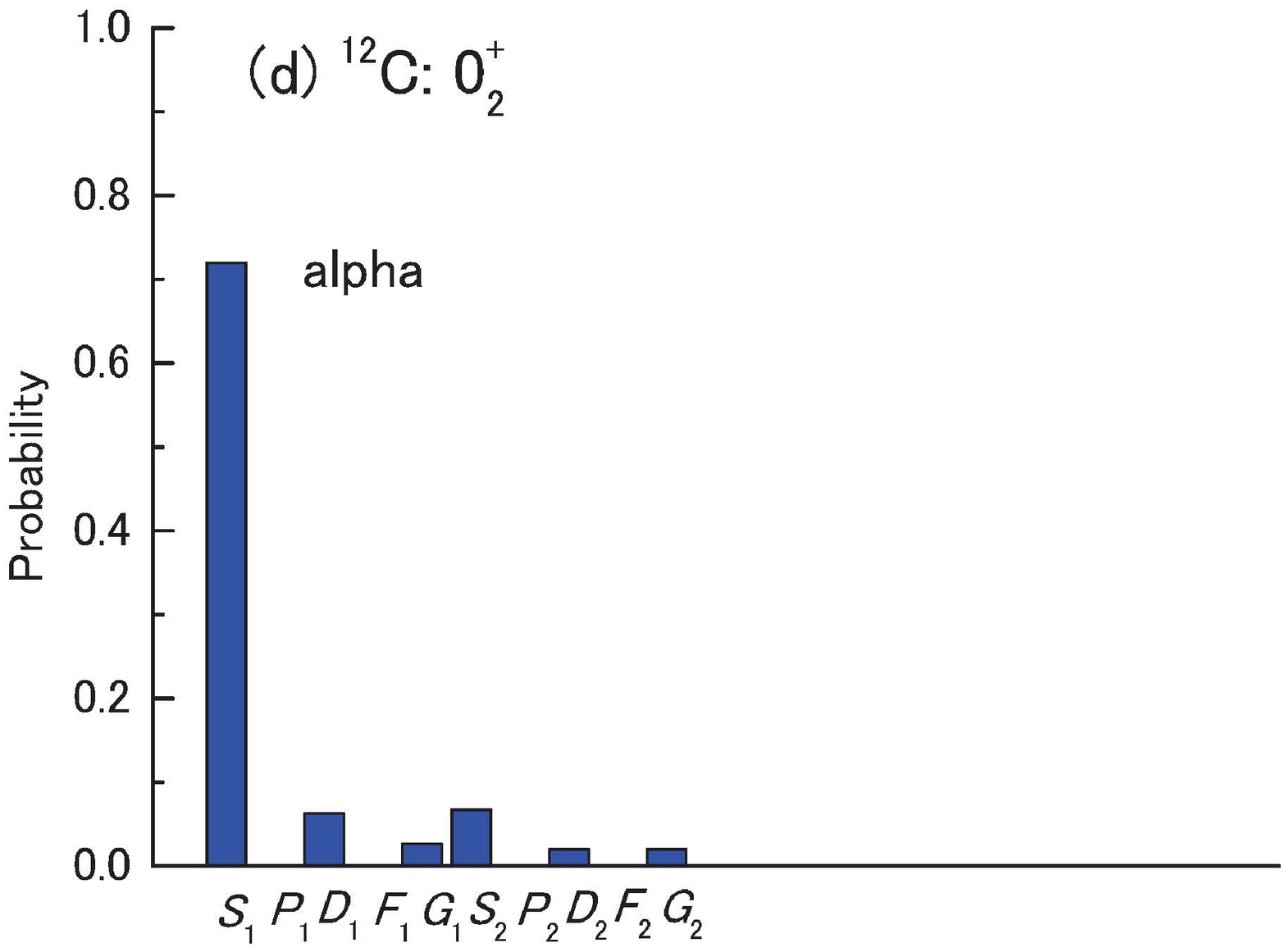,width=\hsize}
\end{center}
\end{minipage}
\end{tabular}
\begin{center}
\caption{
Occupation probabilities of the single-$\alpha$-particle and
 single-$t$-particle orbits for the (a)~$3/2^{-}_{1}$ (g.s.) and (b)~$3/2^{-}_{3}$ states in $^{11}$B.
For reference, we show
 those of the single-$\alpha$-particle orbits for the (c)~$0^{+}_{1}$ (g.s.) and
 (d)~$0^{+}_{2}$ (Hoyle) states in $^{12}$C calculated by the $3\alpha$ OCM~\cite{yamada05}.
}
\label{fig:4}
\end{center}
\end{figure}
%%%%%%%%%%%%%%%%%%%%%%%%%%%%%%%%%%%%%%%%%%%%%%%%%%%

\begin{figure}[tbh]
\begin{tabular}{cc}
\begin{minipage}{0.45\hsize}
\begin{center}
\epsfig{file=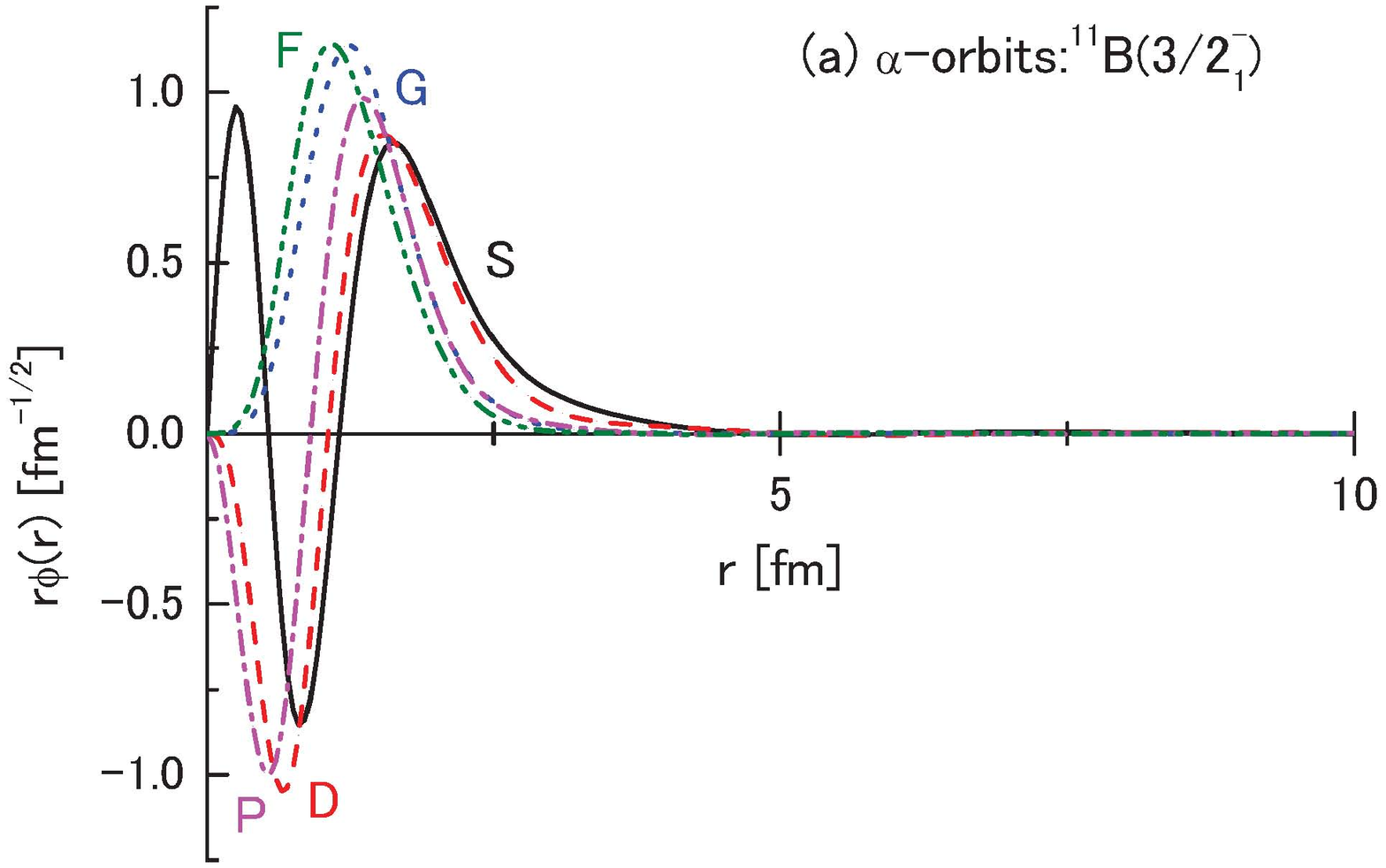,width=\hsize}
\end{center}
\end{minipage}
\hspace*{2mm}
\begin{minipage}{0.45\hsize}
\begin{center}
\epsfig{file=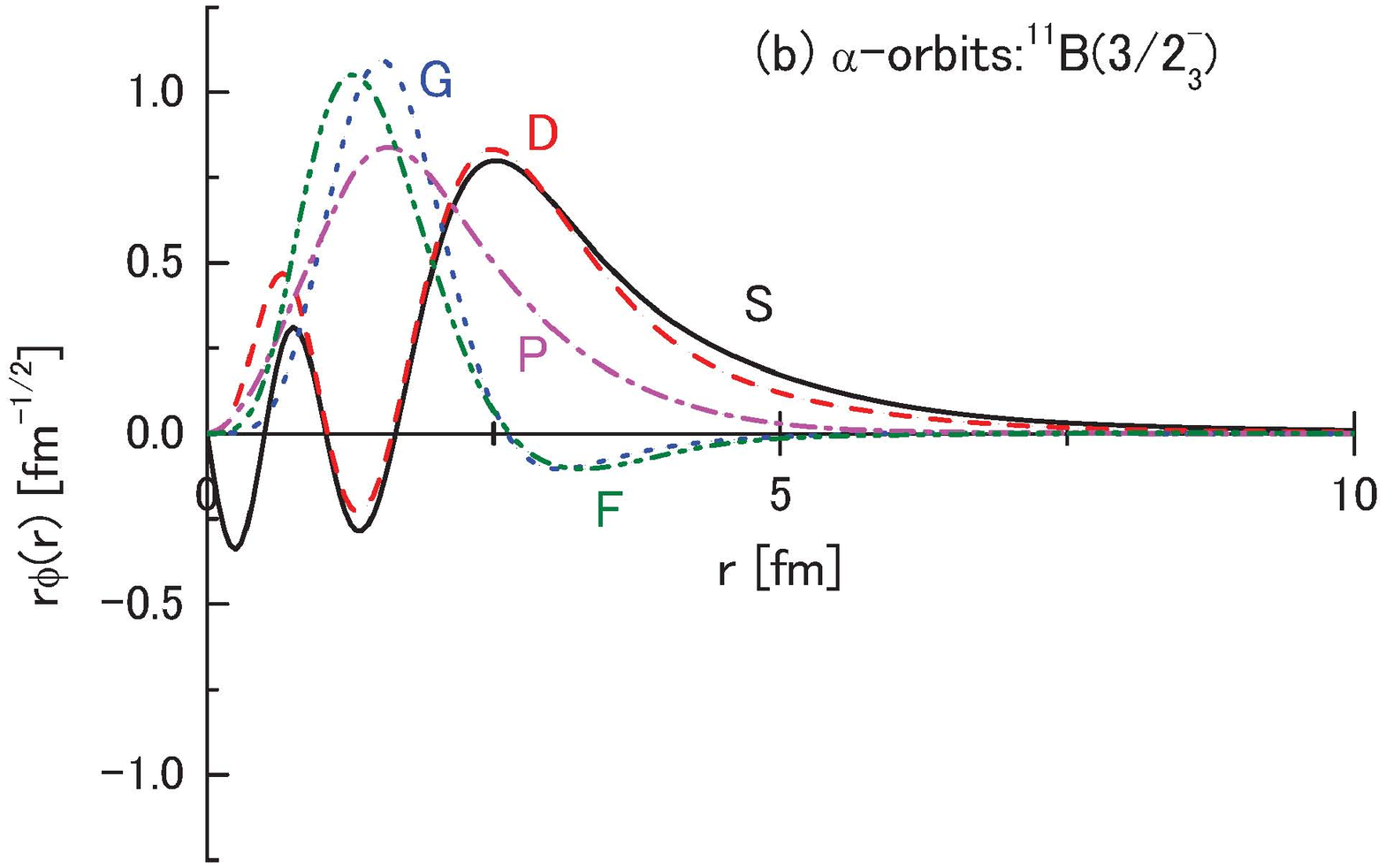,width=\hsize}
\end{center}
\end{minipage}
\\\\
\begin{minipage}{0.45\hsize}
\begin{center}
\epsfig{file=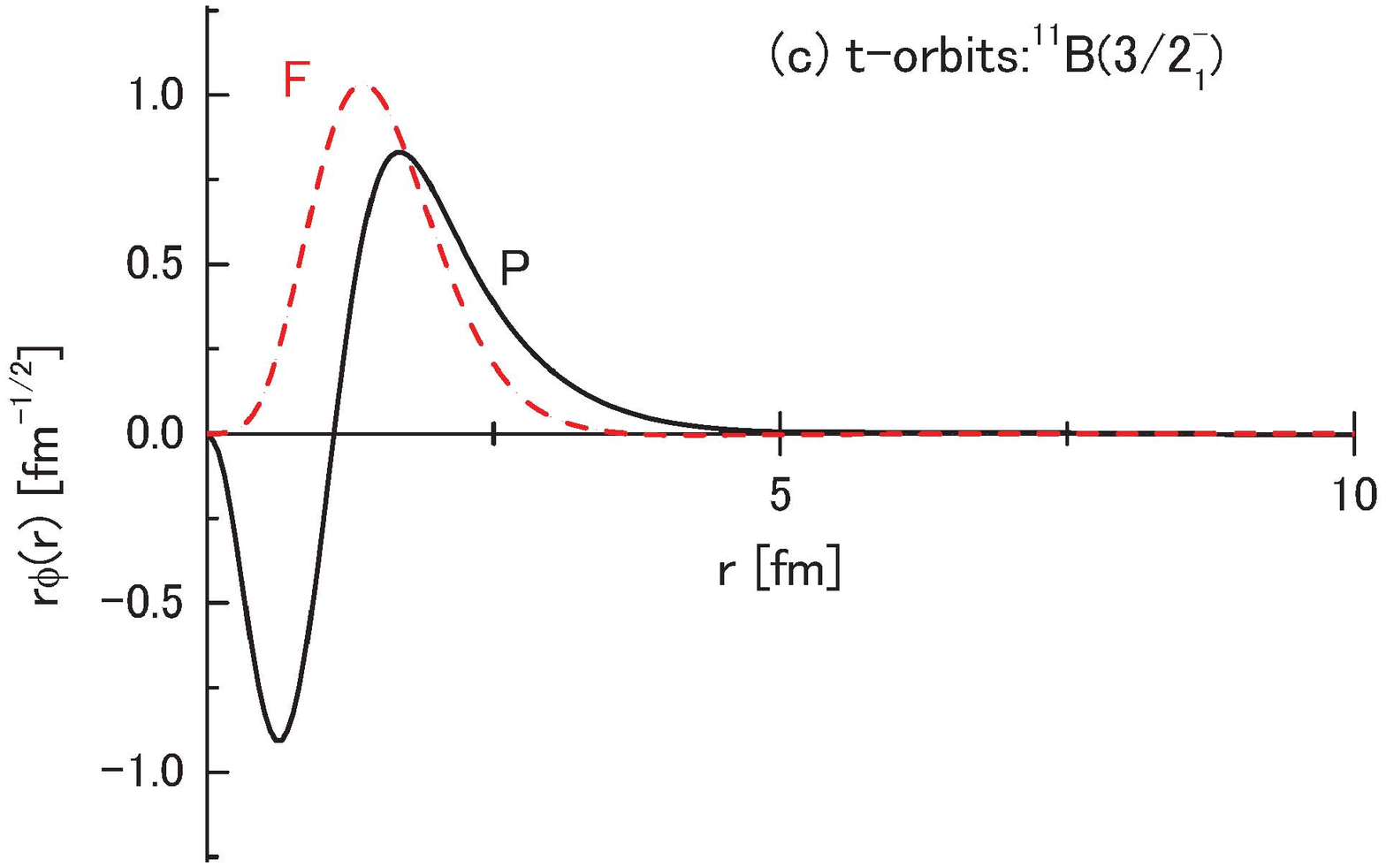,width=\hsize}
\end{center}
\end{minipage}
\hspace*{2mm}
\begin{minipage}{0.45\hsize}
\begin{center}
\epsfig{file=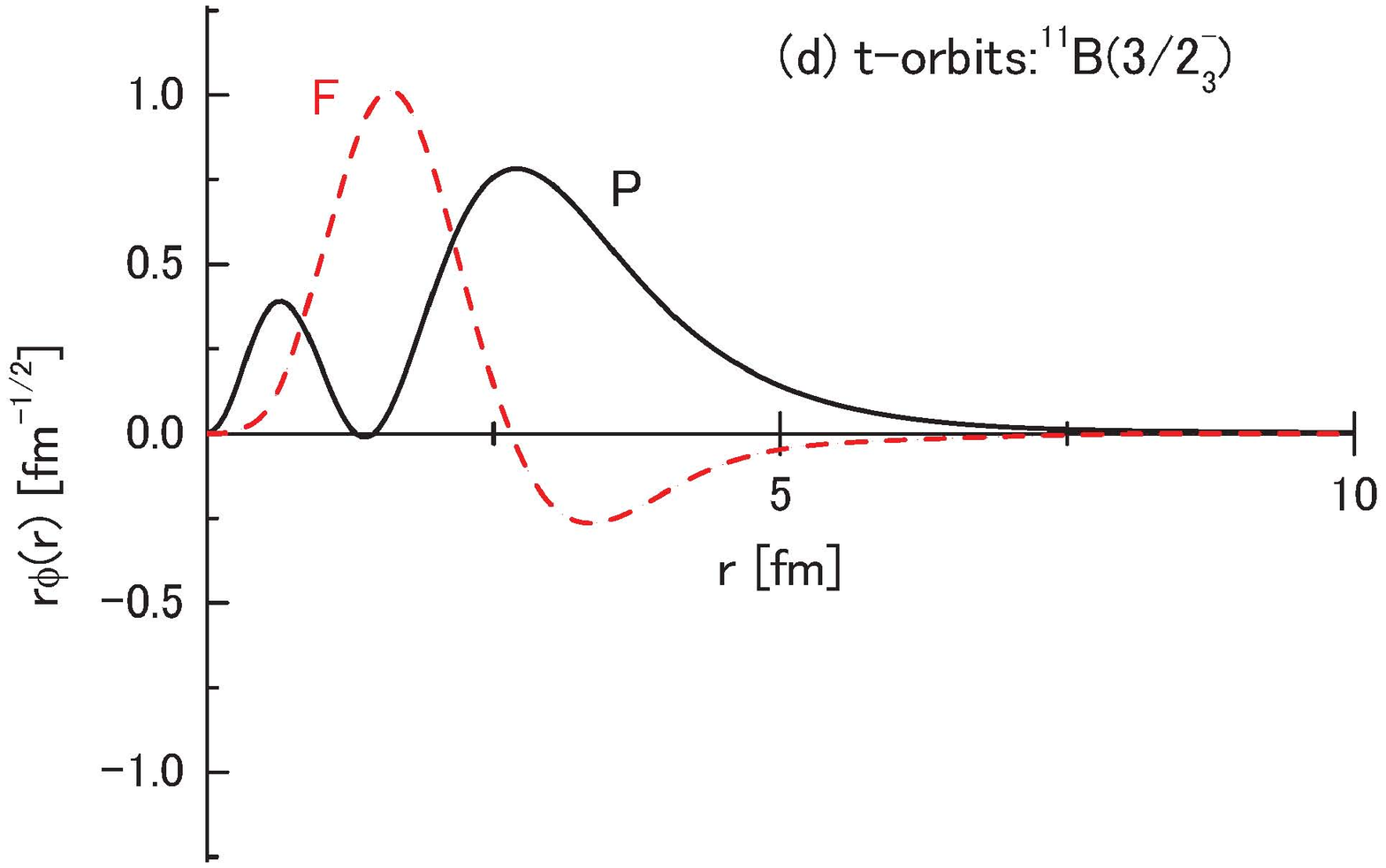,width=\hsize}
\end{center}
\end{minipage}
\\\\
\begin{minipage}{0.45\hsize}
\begin{center}
\epsfig{file=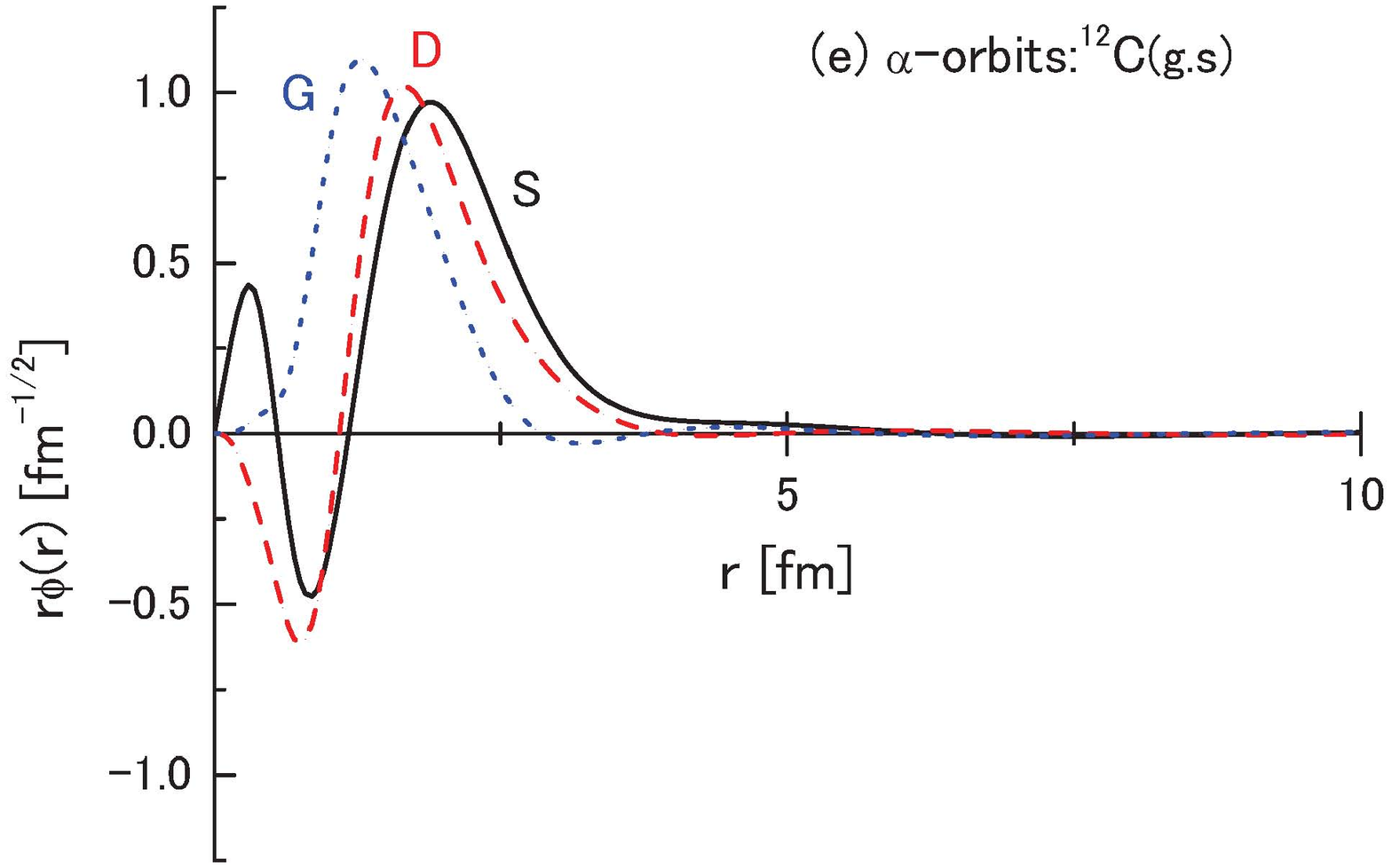,width=\hsize}
\end{center}
\end{minipage}
\hspace*{2mm}
\begin{minipage}{0.45\hsize}
\begin{center}
\epsfig{file=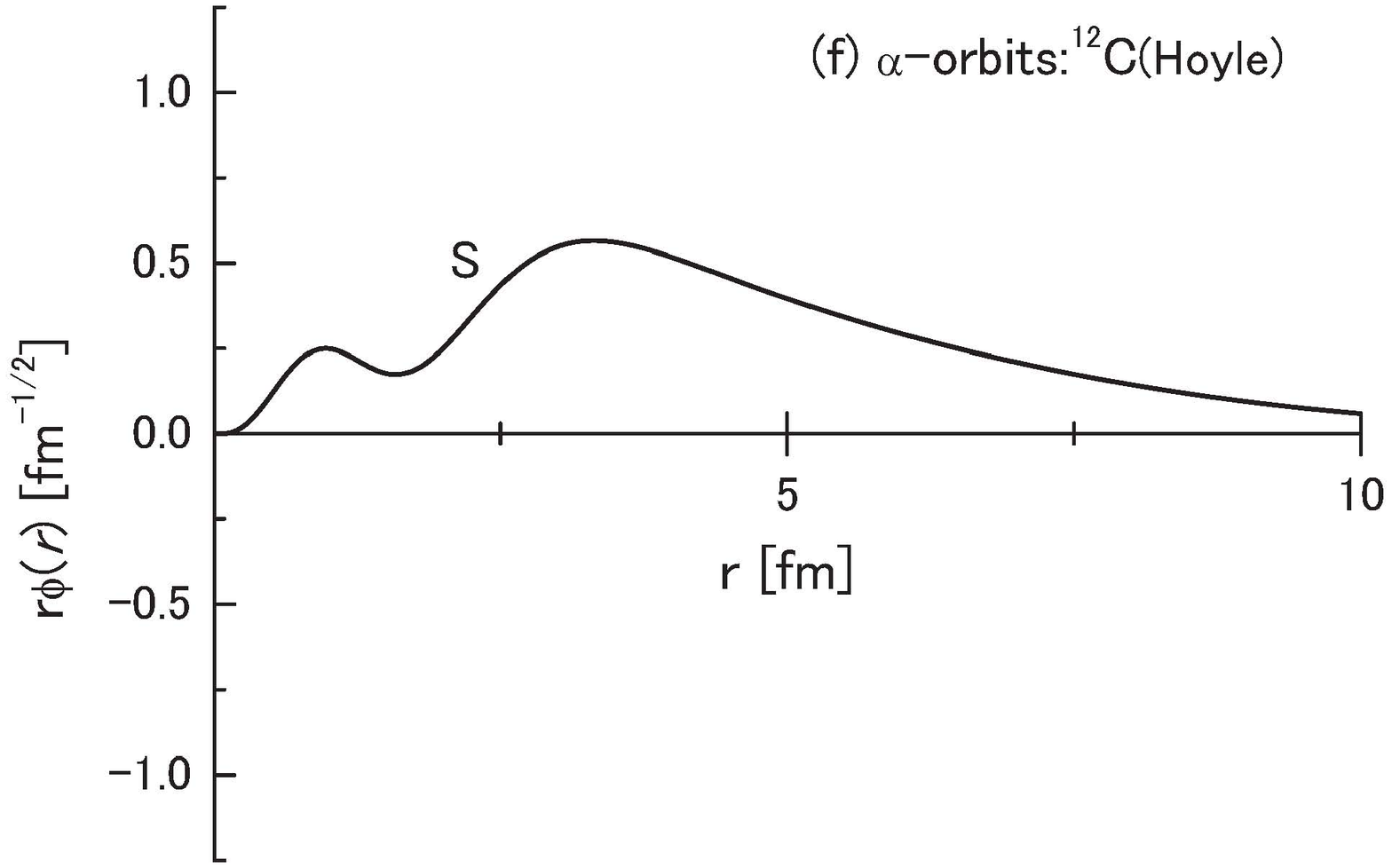,width=\hsize}
\end{center}
\end{minipage}
\end{tabular}
\begin{center}
\caption{
Radial behaviors of the single-$\alpha$-particle orbits in the
 (a)~$3/2^-_1$ and (b)~$3/2^-_3$ states, where
 the solid line (dashed, dotted, dash-dotted, and dash-two-dotted) denotes
 the $S_1$ wave ($D_1$, $G_1$, $P_1$, and $F_1$ respectively),
 and those of the single-$t$-particle orbits
 in the (c)~$3/2^-_1$ and (d)~$3/2^-_3$ states, where
 the solid line (dashed) denotes
 the $P_1$-wave ($F_1$-one)
 [see Figs.~\ref{fig:4}(a) and (b)].
For reference, those of the single-$\alpha$-particle
 orbits in the (e)~ground state of $^{12}$C and
 (f)~Hoyle state are shown~\cite{yamada05},
 where the solid line (dashed and dotted) denotes
 the $S_1$ wave ($D_1$ and $G_1$, respectively). 
 [see Figs.~\ref{fig:4}(a) and (b)].
}
\label{fig:5}
\end{center}
\end{figure}

Figure~\ref{fig:4} shows the occupation probabilities of the {\it n}th 
 $L$-wave single-$\alpha$-particle (single-$t$-particle) orbits, $P^{(\alpha)}_{n}(L)$ ($P^{(t)}_{n}(L)$), 
 in the $3/2^-_1$ and $3/2^-_3$ states. 
In the $3/2^-_1$ state, the occupation probabilities of the single-$\alpha$-particle orbits
 spread out in several orbits:~$P^{(\alpha)}(S_1)=22.3\%$, 
 $P^{(\alpha)}(P_1)=11.1\%$, $P^{(\alpha)}(D_1)=27.6\%$, $P^{(\alpha)}(F_1)=9.6\%$, and
 $P^{(\alpha)}(G_1)=26.8\%$.
The remaining probability, $2.6\%$, belongs to the higher orbits $L_{n}$ with $n\not=1$. 
These features can be understood from the fact that the $3/2^-_1$ state has the dominant configuration
 of SU(3)$[f](\lambda,\mu)_{L}=[443](1,3)_{1}$ ($Q$=7) coupled with the spin of $t$-cluster. 
In fact, the SU(3) wave function is expressed in terms of the harmonic oscillator basis
 $| (n_{44}\ell_{44})\otimes(n_{83}\ell_{83})\rangle$,
\begin{eqnarray}
{| [443](1,3)_{L=1} \rangle} &=& \sqrt{\frac{64}{225}}{| (2S)\otimes(1P)\rangle}
                                - \sqrt{\frac{56}{225}}{| (1D)\otimes(1P)\rangle} \nonumber\\
                                &-& \sqrt{\frac{24}{225}}{| (1D)\otimes(0F)\rangle}
                                + \sqrt{\frac{81}{225}}{| (0G)\otimes(0F)\rangle},
\label{eq:SU3_11B_L1}
\end{eqnarray}
where $(n_{44}\ell_{44})$ [$(n_{83}\ell_{83})$] represents
 the harmonic oscillator wave function, $n$ and $\ell$ denoting the number of nodes
 and the orbital angular momentum, respectively,
 with respect to the relative coordinate $\Vec{r}_{44}$ between the two $\alpha$ clusters
 ($\Vec{r}_{83}$ between the $2\alpha$ and $t$ clusters). 
From the definition of the single-$\alpha$-particle density matrix in Eq.~(\ref{eq:density_alpha}),
 one obtains the occupation probabilities $P^{(\alpha)}(L)$ as follows:~$P^{(\alpha)}(S_1)=22.3\%$, 
 $P^{(\alpha)}(P_1)=11.4\%$, $P^{(\alpha)}(D_1)=28.0\%$, $P^{(\alpha)}(F_1)=10.0\%$, and
 $P^{(\alpha)}(G_1)=28.3\%$.
These results agree well with those obtained by diagonalizing the single-$\alpha$-particle
 density matrix as mentioned above (see also sFig.~\ref{fig:4}). 
The radial behaviors of the $S_1$-, $D_1$- and $G_1$-wave single-$\alpha$-particle orbits
 as well as $P_1$ and $F_1$ orbits are shown in Fig.~\ref{fig:5}(a).  
One sees strong oscillations in the inner regions, coming from the strong Pauli-blocking effect,
 reflecting the nature of the compact shell-model-like structure in this state.
These behaviors are quite similar to those of the ground state of $^{12}$C
 shown in Fig.~\ref{fig:5}(e)~\cite{yamada05}, the structure of which is of
 the shell-model-like~\cite{ikeda80,uegaki77,kamimura77,chernykh07}. 
 
On the other hand, the single-$t$-particle occupation probabilities in the $3/2^{-}_1$ state
 concentrate mainly on the two orbits:~$P^{(t)}(P_1)=53.2\%$ and $P^{(t)}(F_1)=44.4\%$
 (see Fig.~\ref{fig:4}(a)).
These results originate from the SU(3) nature of the $3/2^-_1$ state.
In fact, the definition of the single-$t$-particle density matrix in Eq.~(\ref{eq:density_triton})
 gives $P^{(\alpha)}(P)=53.3\%$ and $P^{(\alpha)}(F)=46.7\%$ for the SU(3) wave function
 in Eq.~(\ref{eq:SU3_11B_L1}).
The reason why the single-$t$-particle probability with odd parity is exactly zero in $3/2^-$ 
 is due to the symmetric nature of the total wave function of $^{11}$B in Eq.~(\ref{eq:total_wf}).
In fact, the symmetric character in Eq.~(\ref{eq:total_wf}) with respect to the exchange
 between two $\alpha$ clusters causes even-parity orbital angular momentum
 between the two $\alpha$'s and thus, only odd-parity-wave single-$t$-particle orbits
 are allowed in the negative-parity $^{11}$B state.
This is discussed in details in the case of $^{12}$C with the $3\alpha$ OCM (see Ref.~\cite{yamada05}).
The radial behavior of the $P$-wave single-$t$-particle orbit in Fig.~\ref{fig:5}(c) indicates
 a strong oscillation in the inner region, reflecting the compact shell-model-like
 feature of the $3/2^-_1$ state.
 
%%%%%%%%%%%% 
%  Fig. 5  %
%%%%%%%%%%%%
%%%%%%%%%%%%%%%%%%%%%%%%%%%%%%%%%%%%%%%%%%

As for the $3/2^{-}_{3}$ state possessing an $\alpha+\alpha+t$ cluster structure,
 the occupation probabilities of the single-$\alpha$-particle orbits have
 {\it no} concentration on a single orbit:~$P^{(\alpha)}(S_1)=28.6\%$, 
 $P^{(\alpha)}(P_1)=10.9\%$, $P^{(\alpha)}(D_1)=28.3\%$, $P^{(\alpha)}(F_1)=4.8\%$,
 and $P^{(\alpha)}(G_1)=9.1\%$ (see Fig.~\ref{fig:4}). 
The remaining probability, $18.3\%$, belongs to the higher orbits $L_{n}$ with $n\not=1$. 
One notices that $P^{(\alpha)}(S_1) \sim P^{(\alpha)}(D_1) \sim 30\%$.
It is noted that there is a close relation between the occupation probability $P^{(\alpha)}$
 and the $S^{2}$ factor of the $\alpha$ reduced width amplitudes, as discussed in Ref.~\cite{yamada05}.
In fact, in the present calculation, the $S$-wave and $D$-wave $S^2$ factors
 of the $^7$Li+$\alpha$ reduced width amplitudes in $3/2^{-}_{3}$ amount to $0.26$ and $0.29$, 
 respectively, the ratio of which is similar to $P^{(\alpha)}(S_1)/P^{(\alpha)}(D_1)\sim 1$.  

The radial behaviors of the single-$\alpha$-particle $S_1$, $D_1$ and $G_1$ orbits 
 are shown in Fig.~\ref{fig:5}(b).
Compared with the results of the ground state (see Fig.~\ref{fig:5}(a)), 
 the strong oscillation in the inner region is rather suppressed,
 although the outermost peak is enhanced in the $3/2^{-}_3$ state.
For reference, we show in Fig.~\ref{fig:5}(f) the $S$-wave single-$\alpha$-particle orbit
 in the Hoyle state of $^{12}$C~\cite{yamada05} (which is a typical example of $\alpha$ condensation in nuclei), 
 where the $\alpha$ particle occupies the $S$ orbit with about $70\%$ probability.
In this Hoyle state, one sees no nodal behaviors (but small oscillations) in the inner region together with
 a long tail, and the radial behavior is likely to be of the $0S$-type Gaussian.
This Gaussian behavior and its occupation probability as large as about $70\%$
 are the evidence that the Hoyle state has a $3\alpha$ condensate
 structure~\cite{yamada05,matsumura04,funaki09}.
Comparing them with the case of the $3/2^{-}_{3}$ state in $^{11}$B,
 we find remarkable qualitative differences between them:~1)~the $3/2^{-}_{3}$ state
 has no concentration of occupation probability on a single-$\alpha$-particle orbit
 and 2)~nodal behaviors remain in the inner region in $3/2^{-}_{3}$, 
 indicating rather large Pauli-blocking effect. 
These results show that the $3/2^{-}_{3}$ state does not have $\alpha$-condensate nature
 like the Hoyle state.

Concerning the single-$t$-particle occupation probabilities, 
 the $3/2^{-}_{3}$ state has $P^{(t)}(P_1)=51.8\%$ and $P^{(t)}(F_1)=23.2\%$,
 as shown in Fig.~\ref{fig:4}.
The remaining probability is attributed to higher orbits $L_{n}$ with $n\not=1$.  
The radial behavior of the $P_1$-wave single-$t$-particle orbit is shown in Fig.~\ref{fig:5}(d).
In contrast with the case of the single-$\alpha$-particle orbits,
 one sees that the radial behavior of the $P_1$-wave single-$t$-particle orbit ($P^{(t)}(P_1)=51.8\%$)
 has no nodal behaviors but somewhat large oscillator behaviors in the inner region.
This radial behavior is likely to be of the $0P$-wave-like.
The reason why the nodal behavior disappears is due to the fact that 
 the rms distance between the center-of-mass of $2\alpha$ and the triton
 is as large as $R_{{^8{\rm Be}}-t}\simeq3.4$~fm (see Table~\ref{tab:1}),
 and thus the Pauli blocking effect is weakened significantly to make the small
 oscillation in the inner region. 

From the above-mentioned results, the $3/2_{3}^{-}$ state could not be identified
 as the analogue of the Hoyle state, although its state has the $\alpha+\alpha+t$ cluster structure.
The essential reason why the $3/2_{3}^{-}$ state is not of the Hoyle-analogue
 is due to the fact that the $3/2^{-}_{3}$ state is bound
 by $2.9$ MeV with respect to the $\alpha+\alpha+t$ threshold, while the
 Hoyle state appears by $0.38$~MeV above the $3\alpha$ threshold,
 i.e.~the energy level is located around the Coulomb barrier produced by the clusters.
This extra binding energy of the $3/2_{3}^{-}$ state with respect to the three-body threshold
 hinders significantly the development of the gas-like $\alpha+\alpha+t$ structure with a large
 nuclear radius in the state.
In Refs.~\cite{yamada04,funaki09}, they discussed important requirements
 for the emergence of the Hoyle-analogue states composing of
 clusters with a large radius, in which all the clusters occupy an $0S$ orbit, 
 and thus its dominant configuration is $(0S)^n$,  where $n$ denotes the number of
 the clusters:~1)~The state should be located above
 the constituent cluster threshold to make a dilute cluster structure with a large nuclear radius, 
 i.e.~it should appear around the Coulomb barrier produced by the relevant constituent clusters, 
 and 2)~all of the clusters interact dominantly in $S$-wave each other to avoid
 the Pauli-blocking effect among the clusters as well as reducing the effect of the centrifugal
 barrier among them.
They are {\it the necessary conditions for the appearance of the Hoyle-analogue states}.
These conditions are not fulfilled in the $3/2_{3}^{-}$ state as discussed in the present study.
Thus, the $3/2_{3}^{-}$ state does not have Hoyle-analogue nature.

%%%%%%%%%%%%%%%%%%%%%%%%%%%%%%%%%
\subsection{$1/2^{+}$ states}
%%%%%%%%%%%%%%%%%%%%%%%%%%%%%%%%%
 
\begin{figure}[tbh]
\begin{center}
\epsfig{file=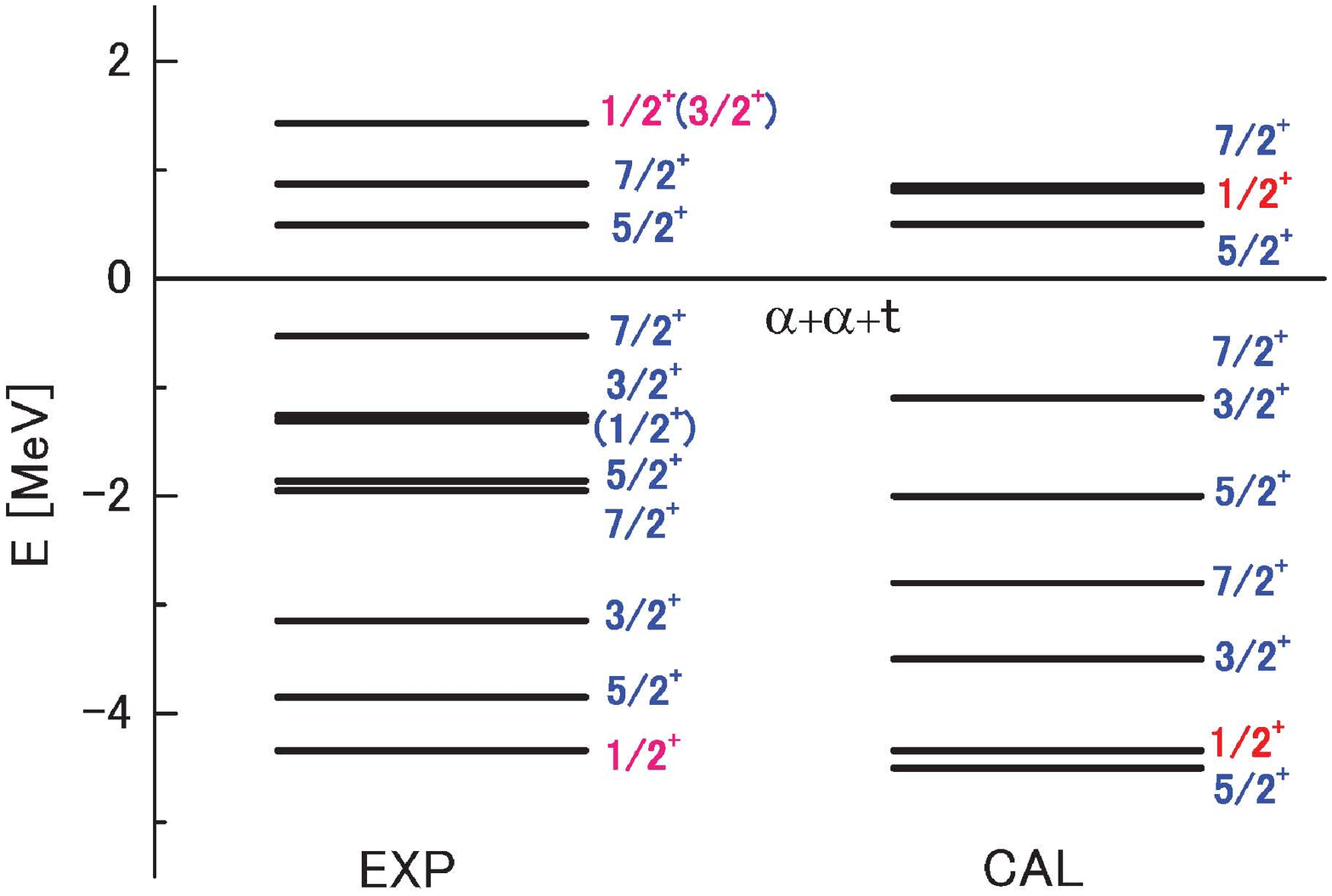,width=0.6\hsize}
\caption{
Calculated energy levels of the positive-parity states in $^{11}$B 
 with respect to the $\alpha+\alpha+t$ threshold together
 with the experimental data~\cite{ajzenberg86}.
}
\label{fig:6}
\end{center}
\end{figure}

The $1/2_{1}^{+}$ state experimentally appears at $E_x=6.9$~MeV (see Fig.~\ref{fig:6}).
The lowest particle decay threshold is the $^{7}$Li+$\alpha$ channel ($E_x=8.7$~MeV)
 which is located at $2.5$ MeV below the $\alpha+\alpha+t$ threshold. 
Since the $1/2_{1}^{+}$ state appears around the $^{7}$Li+$\alpha$ threshold, 
 its structure may have a $^{7}$Li+$\alpha$ cluster structure.
In addition, the small excitation energy of $1/2_{1}^{+}$
 in comparison with $\hbar\omega=41/A^{1/3}\simeq 18$~MeV in this mass region
 suggests that an $\alpha$-cluster correlation should play an important role
 in lowering the excitation energy.
 
The positive-parity levels of $^{11}$B obtained by the present $\alpha+\alpha+t$
 OCM calculation are shown in Fig.~\ref{fig:6} (only $1/2^{+}$ states as well as $3/2^{-}$ states 
 are given in Fig.~\ref{fig:2}).
The $1/2_{1}^{+}$, $3/2_{1}^{+}$ and $5/2_{1}^{+}$ states are stable against any particle decays.
Other calculated states correspond to resonant states which are identified
 with use of CSM, as explained in Sec.~\ref{sec:CSM}.
One can see rather good correspondence to the experimental data.  
In the present paper, we concentrate on studying the structure of $1/2^{+}$ states.
The structures of other positive-parity states will be discussed elsewhere,
 although we found that some even-parity states shown in Fig.~\ref{fig:6} have
 cluster structures.  

%%%%%%%%%%%% 
%  Fig. 6  %
%%%%%%%%%%%%
%%%%%%%%%%%%%%%%%%%%%%%%%%%%%%%%%%%%%%%%%%%%%%%%%  

The calculated nuclear radius of $1/2_{1}^{+}$ is $R_{N}=2.82$~fm.
This value is similar to the $3/2_{3}^{-}$ state which has the $\alpha+\alpha+t$
 cluster structure (see Table~\ref{tab:1}).
Figure~\ref{fig:3}(a) shows the reduced width amplitude or overlap amplitude for
 the $^7$Li($3/2^{-}$:g.s)+$\alpha$ channel with the relative orbital angular
 momentum $\ell_{74}=1$.
One sees that the shape of the outermost peak at $r\sim5$ fm is similar to that of $3/2_{3}^{-}$.
The $S^2$ factor for this channel is $0.203$.
This is the largest among all of the $^7$Li+$\alpha$ channels
 with $^{7}$Li$({\rm g.s},1/2^-,7/2^-,5/2^-)$, where the corresponding $S^2$ factors are $0.203$, $0.099$,
 $0.120$, and $0.047$, respectively.
As for the $^8$Be+$t$ channel, the $S^2$ factor for $^8$Be(g.s)+$t$
 [$^8$Be($2^+$)+$t$] is $0.130$ ($0.134$). 
The reduced width amplitude of the $^8$Be(g.s)+$t$ channel is shown in Fig.~\ref{fig:3}(b).
On the other hand the $\alpha - t$ distance ($^7{\rm Li}-\alpha$) in $1/2_{1}^{+}$ is 
 $R_{\alpha-t}=3.81$ fm ($R_{^7{\rm Li}-\alpha}=3.45$ fm).
This value is a little bit larger than that of the $^7$Li ground state 
 ($R_{\alpha-t}=3.5$ fm) obtained by the $\alpha + t$ OCM calculation.
Thus, the $1/2_{1}^{+}$ state has a dominant structure of $^7$Li(g.s)+$\alpha$
 with $P$-wave relative motion.
As discussed in Sec.~\ref{sec:3/2-}, the $3/2^{-}_{3}$ state has the largest $S^{2}$ factor
 for the $^{7}$Li(g.s)+$\alpha$ channel with $S$-wave relative motion.
This means that the state includes significantly the structure of $^{7}$Li(g.s)+$\alpha$. 
In the non-identical two-cluster states, for example, in the $^{16}$O+$\alpha$ states of $^{20}$Ne,
 there appear the parity-doublet states, $0^{+}$ and $1^-$, 
 around the $^{16}$O+$\alpha$ threshold, in which the relative motion between
 $^{16}$O and $\alpha$ is of the $S$-wave and $P$-wave, respectively~\cite{ikeda80,horiuchi68}. 
Thus, the $1/2_{1}^{+}$ and $3/2^{-}_{3}$ states of $^{11}$B, which are located around
 the $^{7}$Li(g.s)+$\alpha$ threshold, can be interpreted as the parity doublet in $^{11}$B.

%%%%%%%%%%%% 
%  Fig. 7  %
%%%%%%%%%%%%
\begin{figure}[tbh]
\begin{tabular}{cc}
\begin{minipage}{0.45\hsize}
\begin{center}
\epsfig{file=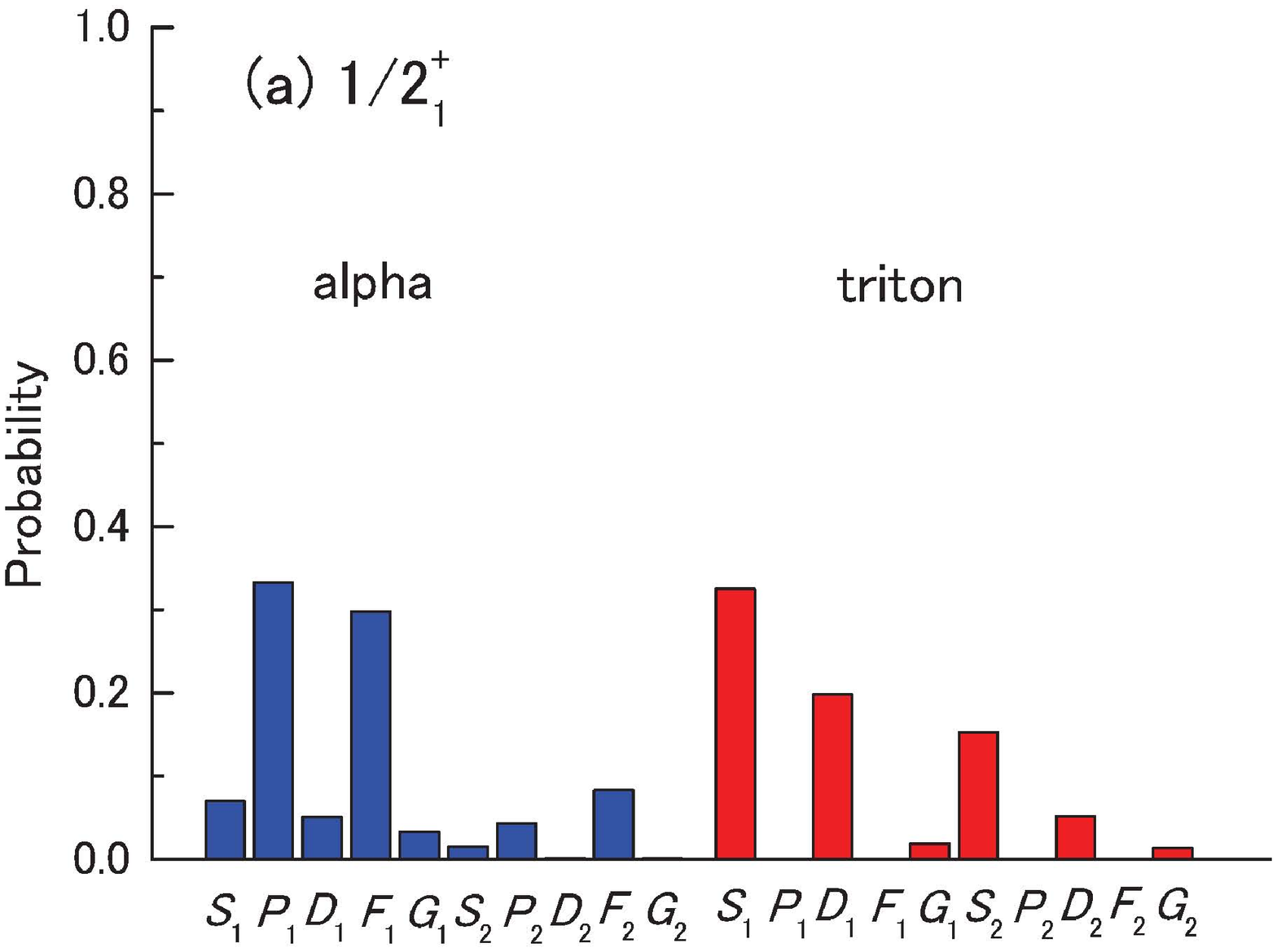,width=\hsize}
\end{center}
\end{minipage}
\hspace*{2mm}
\begin{minipage}{0.45\hsize}
\begin{center}
\epsfig{file=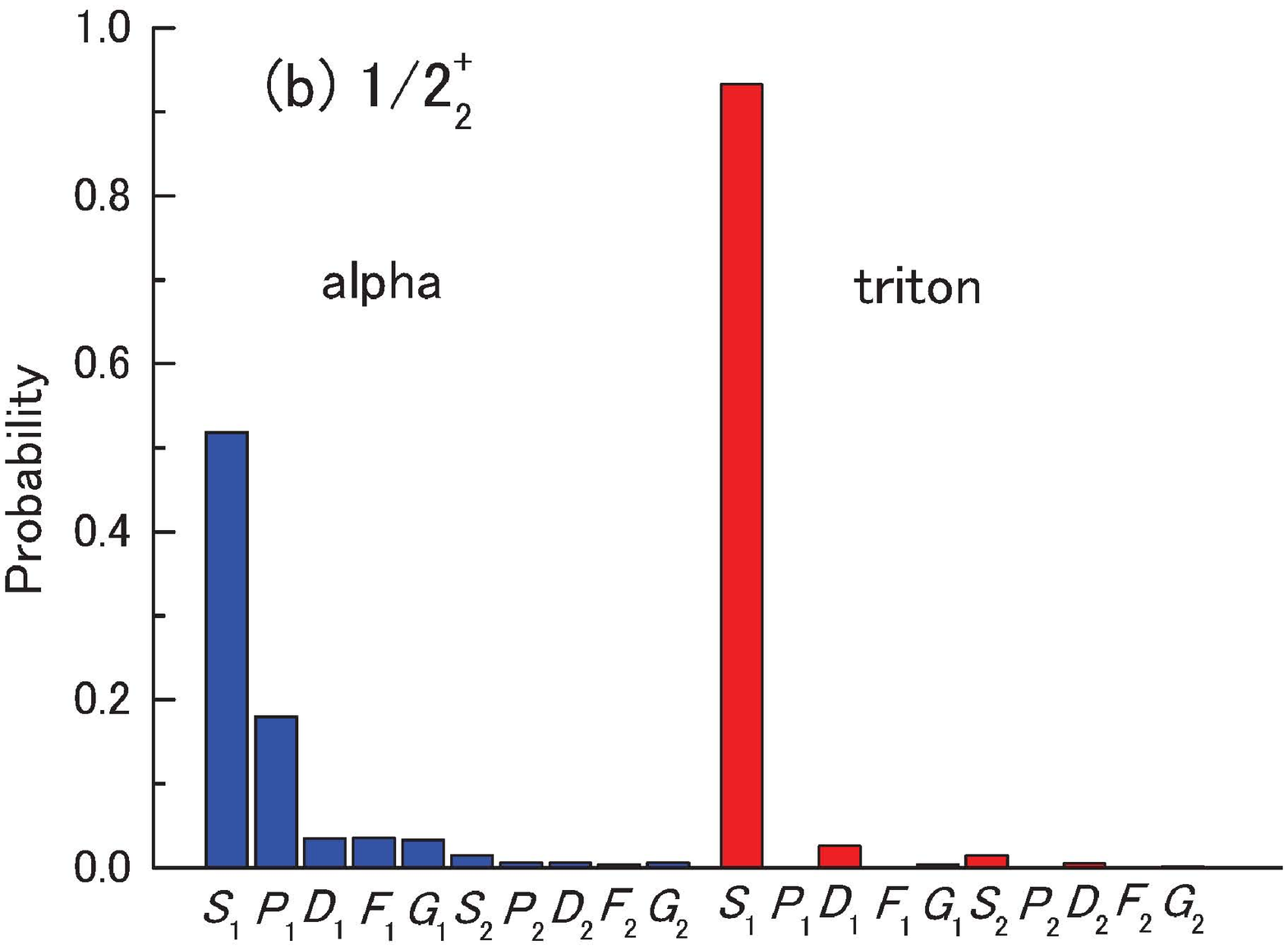,width=\hsize}
\end{center}
\end{minipage}
\end{tabular}
\begin{center}
\caption{
Occupation probabilities of the single-$\alpha$-particle and
 single-$t$-particle orbits for the (a)~$1/2^{+}_{1}$  and (b)~$1/2^{+}_{2}$ states in $^{11}$B.
}
\label{fig:7}
\end{center}
\end{figure}
%%%%%%%%%%%%%%%%%%%%%%%%%%%%%%%%%%%%%%%%%%%%%%%%%%%

\begin{figure}[tbh]
\begin{tabular}{cc}
\begin{minipage}{0.45\hsize}
\begin{center}
\epsfig{file=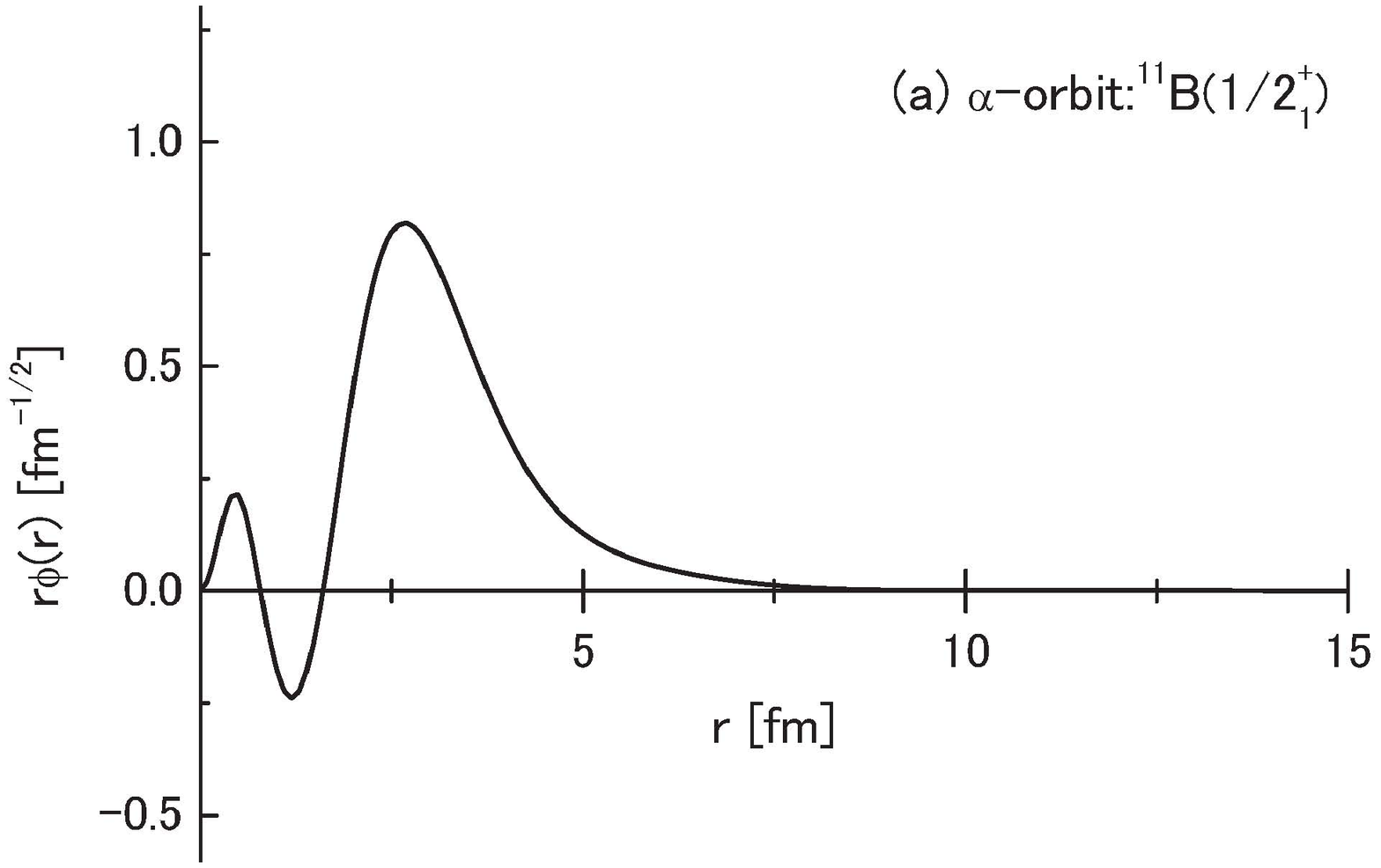,width=\hsize}
\end{center}
\end{minipage}
\hspace*{2mm}
\begin{minipage}{0.45\hsize}
\begin{center}
\epsfig{file=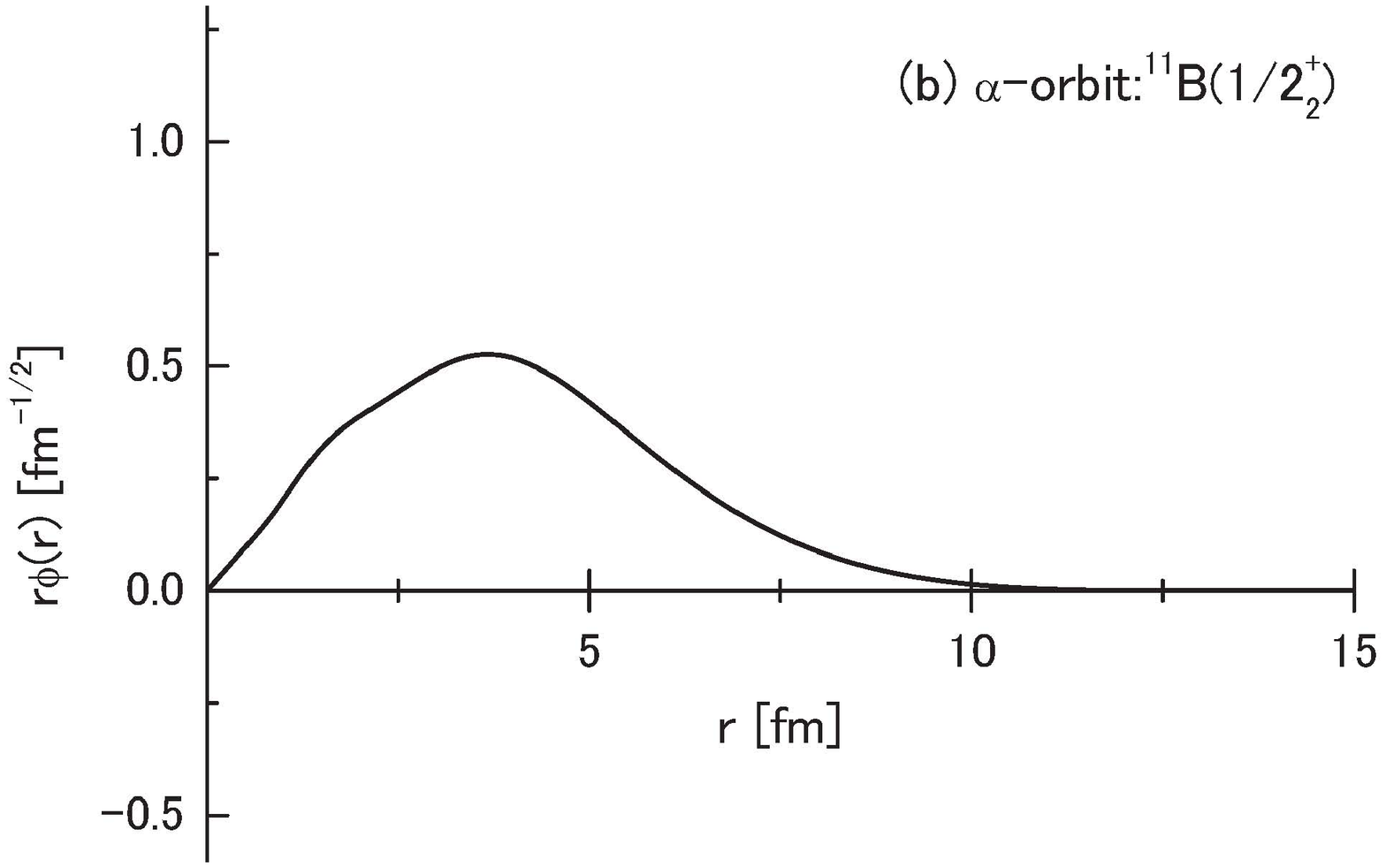,width=\hsize}
\end{center}
\end{minipage}
\\\\
\begin{minipage}{0.45\hsize}
\begin{center}
\epsfig{file=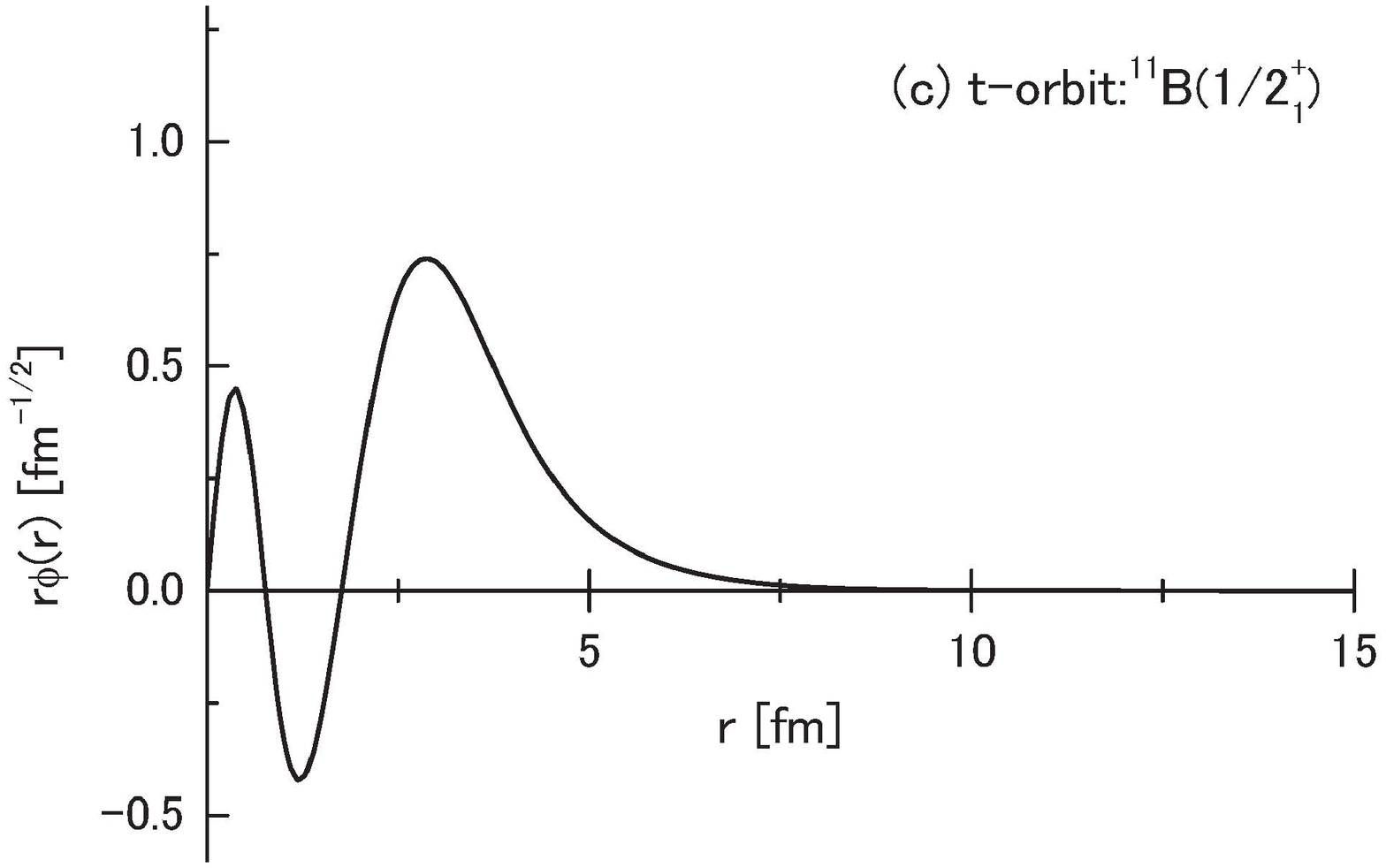,width=\hsize}
\end{center}
\end{minipage}
\hspace*{2mm}
\begin{minipage}{0.45\hsize}
\begin{center}
\epsfig{file=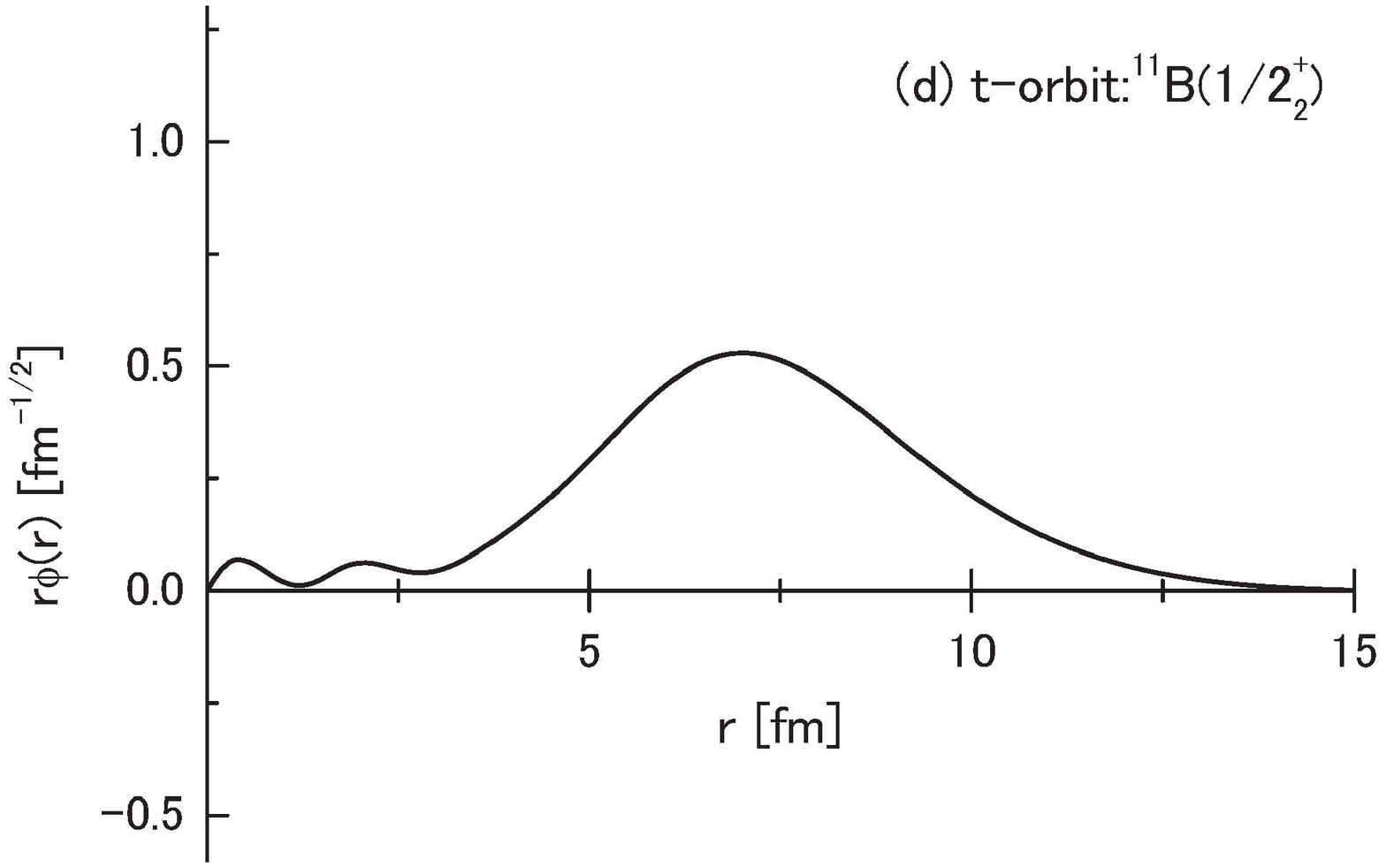,width=\hsize}
\end{center}
\end{minipage}
\end{tabular}
\begin{center}
\caption{
Radial behaviors of the single-$\alpha$-particle orbits in the $1/2_1^+$
 and $1/2_2^+$ states:~(a)~$P_1$-wave of $1/2^+_1$  
 with the largest occupation probability $P^{(\alpha)}(P_1)=33$~\%
 and (b)~$S_1$-wave of $1/2^+_2$ with $P^{(\alpha)}(S_1)=52$~\% 
 [see Figs.~\ref{fig:7}(a) and (b)].
Those of the single-$t$-particle orbits in the two $1/2$ states are shown
 in (c)~for the $S_1$-wave of $1/2^+_1$
 with the largest occupation probability $P^{(t)}(S_1)=33$~\%
 and in (d) for $S_1$-wave of  $1/2^+_2$ with $P^{(t)}(S_1)=93$~\%.
}
\label{fig:8}
\end{center}
\end{figure}

The occupation probabilities of the single-$\alpha$-particle and single-$t$-particle
 orbits for $1/2_{1}^{+}$ obtained by diagonalizing the single-cluster density matrices
 [see Eqs.~(\ref{eq:single_alpha_density}) and (\ref{eq:single_triton_density})]
 are shown in Fig.~\ref{fig:7}(a).
One sees that the $P_1$ ($F_1$) orbit of $\alpha$-particle has the largest
 (secondarily largest) occupation probability  of $33$~\% ($30$~\%).
A close relationship between the cluster occupation probabilities and $S^2$ factors
 (as well as one between the single-cluster orbits and reduced width amplitudes)
 is discussed in detail in Refs.~\cite{yamada05,funaki08_16O}.
In fact, one can see the close relations in the present results:~1)~the $S^2$ factor
 of $^7$Li(g.s)+$\alpha$  with $\ell_{74}=1$ ($\ell_{74}=3$) is the largest (secondarily largest)
 among all of the $^7$Li+$\alpha$ channels, as discussed above, 
 and 2)~the radial behavior of the $P_1$ orbit in Fig.~\ref{fig:8}(a) is similar
 to that of the reduced width amplitude the $^7$Li(g.s)+$\alpha$ channel in Fig.~\ref{fig:3}(a).
It should be reminded that the occupation probabilities satisfy with a sum rule, while the
 $S^2$ factors do not. 
On the other hand, the occupation probabilities of single-$t$-particle orbits spread out over
 $S$-, $D$-, and $G$-orbits.
This feature is similar to the results of the $S^2$ factors for the $^8$Be+$t$ channels.
The reason why the odd-parity $t$-orbit occupation probabilities are exactly zero
 comes from a symmetric feature of the $\alpha+\alpha+t$ cluster model
 with respect to the exchange of the two $\alpha$'s, as mentioned above
 (see also Ref.~\cite{yamada05}). 
The present results indicate that the $1/2_{1}^{+}$ state is not of the Hoyle-analogue, 
 in which all clusters dominantly occupy the $0S_{\alpha}$ orbit.
The main reason why the $1/2_{1}^{+}$ state does no have a dilute structure like the Hoyle state is
 due to the fact that $1/2_{1}^{+}$ is bound by 4.2 MeV with respect to the $\alpha+\alpha+t$
 threshold. 
This fact hinders strongly the growth of a gas-like $\alpha+\alpha+t$ structure in $1/2_{1}^{+}$.

%%%%%%%%%%%% 
%  Fig. 7_1  %
%%%%%%%%%%%%
\begin{figure}[tbh]
\begin{center}
\epsfig{file=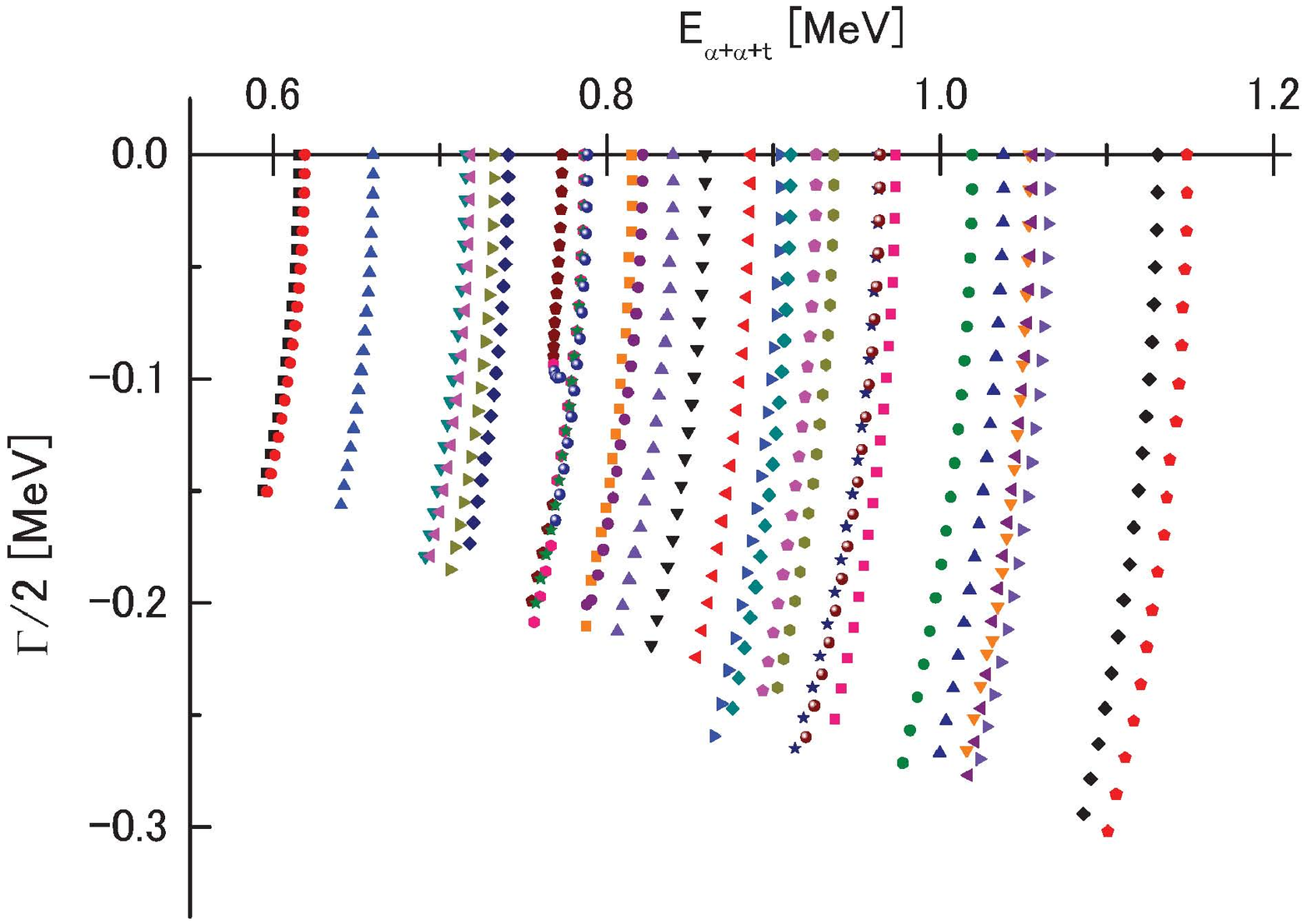,width=0.6\hsize}
\caption{
Trajectory of the energy eigenevalues of $1/2^{+}$ states obtained by solving the the complex-scaled
 Schr\"odinger equation in Eq.~(\ref{eq:CSM})
 as a function of the scaled angle $\theta$
 ($0^{\circ} \leq \theta \leq 9^{\circ }$ with the step of $0.5^{\circ }$).
The real energy $E_{\alpha+\alpha+t}$ is measured from the $\alpha+\alpha+t$ threshold.
In this figure, only the energy region of $0.6~{\rm MeV} \leq E_{\alpha+\alpha+t} \leq 1.2~{\rm MeV}$
 is shown.
A stationary point at $E=(0.75,-0.095)$ gives the resonance parameters. 
Other points correspond to the discretized $\alpha+\alpha+t$ continuum states.
}
\label{fig:complex_scaling}
\end{center}
\end{figure} 
%%%%%%%%%%%%%%%%%%%%%%%%%%%%%%%%%%%%%%%%%%%%%%%%%  

The CSM with the $\alpha+\alpha+t$ OCM is a powerful tool to identify
 resonant states, in particular, those around the $\alpha+\alpha+t$ threshold.
Figure~\ref{fig:complex_scaling} displays the trajectory of the energy eigenevalues obtained by
 solving the complex-scaled Schr\"odinger equation in Eq.~(\ref{eq:CSM}) as a function of the scaled angle $\theta$.
As explained in Sec.~\ref{sec:CSM}, one can identify bound states and resonances as stationary points
 independently of $\theta$. 
In the present study, in addition to the bound $1/2^+_1$ state, 
 we found that the $1/2^{+}_{2}$ state appears at $E_x=11.85$~MeV,
 ($E=0.75$~MeV just above the $\alpha+\alpha+t$ threshold), as a resonant state
 with the width of $\Gamma=190$~keV.
It is reminded that the Hoyle state ($^{12}$C$(0^+_2)$) appears at $0.38$~MeV above the $3\alpha$ threshold,
 and has the very small width of $\Gamma=8.5\pm1.0$~eV. 

%%%%%%%%%%%% 
%  Fig. 8  %
%%%%%%%%%%%%
%%%%%%%%%%%%%%%%%%%%%%%%%%%%%%%%%%%%%%%%%%%%%%%

It is instructive to study the structure of the $1/2_2^+$ state under the bound-state
 approximation, because the calculated width is small.  
We found that the calculated nuclear radius of $1/2_2^+$ is $5.93$ fm, the value of
 which is significantly larger than that of $1/2_1^+$.
This means that the $1/2_2^+$ state has a dilute cluster structure.
In order to study the structure further, we calculated the occupation probabilities and 
 radial behaviors of the single-$\alpha$-particle and single-$t$-particle orbits by diagonalizing
 the single-cluster density matrices defined in Eqs.~(\ref{eq:single_alpha_density})
 and (\ref{eq:single_triton_density}).
The results are shown in Fig.~\ref{fig:7}(b).
We see that the occupation probability of $S_1$-wave $\alpha$-orbit ($S_{\alpha}$)
 is as large as about $52~\%$, and that of $S_1$-wave $t$-orbit ($S_t$)
 amounts to about $93~\%$.
Then the probability of the $(S_\alpha)^2(S_t)$ configuration in $1/2_2^+$ can be
 estimated to be $1.97/3$ $\sim$ 65\%, where the dominator denotes
 the total cluster number (3 clusters) and the numerator is the cluster number occupied
 in the $(0S_\alpha)^2(0S_t)$ configuration, i.e.~$2\times{0.55}+{0.93}=1.97$.
The radial behaviors of the $S_{\alpha}$-orbit and $S_{t}$-orbit are depicted
 in Figs.~\ref{fig:8}(c) and \ref{fig:8}(d), respectively, 
 showing a Gaussian behavior with no nodes.
Thus, we could call this state the Hoyle-analogue, where all clusters
 are mainly in their respective $0S$ orbits, i.e.~$(0S_\alpha)^2(0S_t)$
 with the probability of about 65\%,
 which is similar to the Hoyle state with the main configuration of $(0S_\alpha)^3$
 possessing the probability of about $70$\%.  
It is reminded that the present $1/2^{+}_2$ state satisfies the 
 necessary conditions of appearing the Hoyle-analogue states as discussed
 in Sec.~\ref{sec:3/2-}.  
The peak position in the $S_{\alpha}$-orbit ($S_{t}$-orbit) is around 3.5~fm (7.0~fm).
This result means that the triton cluster moves outside of the $\alpha$ clusters.
Thus, the $1/2_2^+$ state has a $t$-halo-like structure around the two $\alpha$ clusters.    
It should be reminded that the $S$-wave $\alpha-\alpha$ interaction is attractive enough
 to produce the resonant state of $^8$Be(g.s), but the $S$-wave $\alpha-t$ interaction
 is weakly attractive and has no ability to produce bound/resonant states in $^7$Li.
This qualitative difference makes the $t$-halo-like structure around the two $\alpha$
 clusters in the $1/2^{+}_2$ state.

Here, it is interesting to discuss the reason why the calculated width of
 the $1/2^+_2$ state is small ($\Gamma^{\rm cal}=190$~keV), in sipte of 
 the large excitation energy ($E_x=11.85$ MeV).
The decay width is composed of the contributions from the two-body decays ($^{7}$Li+$\alpha$
 and $^{8}$Be+$t$) and three-body decay ($\alpha+\alpha+t$).
As for the three-body decay, it is suppressed strongly due to the very small
 phase space arising from the fact that the $1/2^+_2$ state is located
 at $0.75$~MeV above the $\alpha+\alpha+t$ threshold.
On the other hand, the $^8$Be(g.s)+$t$ decay is possible, because its state appears
 at about $0.7$~MeV above the $^8$Be(g.s)+$t$ threshold.
But, its energy level is located inside the Coulomb barrier (no centrifugal barrier for
 the $S$-wave decay of $^8$Be(g.s)+$t$), and thus its decay is hindered.  
As for the $^7$Li+$\alpha$ decay, the energy of $1/2^+_2$ measured
 from the $^7$Li(g.s:$3/2^{-}$)+$\alpha$ threshold is about $3.8$~MeV.
In this case, the orbital angular momentum of the decaying $\alpha$ particle
 should be $P$-wave, but the $P$-wave occupation probability of $\alpha$ particle in the state
 is as small as $20~\%$ (see Fig.~\ref{fig:7}).
Thus, the $P$-wave decay of $^7$Li(g.s)+$\alpha$ is suppressed largely due to
 the small occupation probability and the Coulomb and centrifugal barriers.
On the contrary, the $S$-wave decay of $^7$Li+$\alpha$ is also possible,
 because the $S$-wave occupation probability of $\alpha$ particle is about $53~\%$.
However, the bound and/or resonant states of positive-parity $^7$Li states
 have not been identified experimentally so far in the low-energy region
 up to $E_x\sim15$ MeV~\cite{ajzenberg86}.
This experimental fact indicates a large hindrance of the $S$-wave $^7$Li+$\alpha$ decay
 with even-parity $^7$Li states.  
Consequently the calculated width of the $1/2^+_2$ state becomes small.

It is interesting to discuss the experimental situation on the $1/2^+$ states
 (isospin $T=1/2$) of $^{11}$B around the $\alpha+\alpha+t$ threshold.
As mentioned in Sec.~\ref{sec:intro}, the $1/2^+(3/2^+)$ state
 at $E_{x}=12.56$ MeV with $\Gamma=210\pm20$ keV
 (located at $1.4$ MeV above the $\alpha+\alpha+t$ threshold), 
 which was identified so far as isospin $T=3/2$ state~\cite{ajzenberg86},
 has recently been observed through the $\alpha$+$^7$Li decay
 channel~\cite{soic04,curtis05,charity08}.
Comparing the experimental results with our present theoretical results,
 the energy and width of the $1/2_2^+$ state obtained by the present study
 are in good correspondence to the experimental data.
Thus, the present calculated results indicate that the $12.56$-MeV state could be
 assigned as the $1/2_2^+$ state with the Hoyle-analogue structure.

%%%%%%%%%%%%%%%%%%%%%%%%%%%%%%%%%%%%%%%%%%%%%%%%%%%%%%%%%%%%%%%%%%
\section{Summary}\label{sec:summary}
%%%%%%%%%%%%%%%%%%%%%%%%%%%%%%%%%%%%%%%%%%%%%%%%%%%%%%%%%%%%%%%%%%

We have studied the structure of $3/2^-$ and $1/2^+$ states in $^{11}$B
 with the $\alpha+\alpha+t$ OCM using GEM.
The model space covers to describe $^7$Li+$\alpha$ and $^8$Be+$t$ cluster structures
 and $\alpha+\alpha+t$ gas-like structures including the shell-model-like structures.
Full levels up to $3/2^{-}_{3}$ and $1/2^{+}_{2}$ around the $\alpha+\alpha+t$ threshold
 are reproduced.
The $3/2_{1}^{-}$ (g.s) and $3/2_{2}^{-}$ states are found to have the shell-model-like
 compact structures, the results of which are consistent with the previous cluster
 model analyses by Nishioka et al.
The $3/2_{3}^{-}$ state is characterized by the monopole transition strength
 as large as the $0^+_2$ at $E_x=7.65$ MeV in $^{12}$C.
The present study succeeded in reproducing its excitation energy and monopole transition strength.
We found that this state has an $\alpha+\alpha+t$ cluster structure with the nuclear radius
 of $R_{N}=3.00$ fm.
The study of the single-cluster properties such as single-cluster orbits and
 occupation probabilities for $3/2^-_3$ showed that there is no concentration of
 the single-$\alpha$ occupation probability on a single orbit and the radial part of
 the single-$\alpha$ orbits has nodal behaviors in the inner region,
 illustrating rather strong Pauli-blocking effect.  
These results are in contrast with those of the Hoyle state with the dilute
 $3\alpha$-condensate-like character $(0S_{\alpha})^{3}$, in which
 the $\alpha$ particle occupies a single $0S$-orbit (zero node) with about $70$\% probability.
Consequently the $3/2_{3}^{-}$ state could not be identified as the analogue of
 the Hoyle state, possessing a dominant gas-like configuration of $(0S_{\alpha})^{2}(0S_{t})$.
The reason why the $3/2_{3}^{-}$ state is not of the Hoyle-analogue
 can be understood from the following facts:~The $3/2^{-}_{3}$ state is bound
 by $2.9$ MeV with respect to the $\alpha+\alpha+t$ threshold, while the
 Hoyle state is located by $0.38$~MeV above the $3\alpha$ threshold and
 has a dilute $3\alpha$ structure.
The extra binding energy of the $3/2_{3}^{-}$ state with respect to the
 $\alpha+\alpha+t$ threshold is likely to suppress strongly the growth of
 the gas-like $\alpha+\alpha+t$ structure in this state.

As for the $1/2^+$ states, the $1/2^+_1$ state appears as a bound state
 at $E_{x}^{\rm exp}=6.79$ MeV around $^7$Li+$\alpha$ threshold.
This low excitation energy indicates that $\alpha$-type correlation should
 play an important role in the state.  
In fact, we found that the $1/2^+_1$ state with $R_{N}=3.14$~fm
 has the $^7$Li(g.s)+$\alpha$ structure with $P$-wave relative motion, 
 although the $^7$Li($\alpha+t$) part
 is rather distorted in comparison with the ground state of $^7$Li.
Since the $3/2^{-}_{3}$ state has the largest $S^{2}$ factor
 for the $^{7}$Li(g.s)+$\alpha$ channel with $S$-wave relative motion
 compared with those of other $^{7}$Li+$\alpha$ and $^{8}$Be+$t$ channels,
 the $1/2_{1}^{+}$ and $3/2^{-}_{3}$ states of $^{11}$B can be interpreted as
 the parity-doublet partners each other.
They are similar to the typical example of the parity doublet, $0^+_1$
 and $1^-_1$ states in $^{20}$Ne, with the $^{16}$O+$\alpha$ cluster
 structure~\cite{ikeda80,horiuchi68}.

In addition to $1/2_{1}^{+}$, we found that the $1/2^{+}_{2}$ state appears
 as a resonant state at $E_x=11.95$ MeV ($\Gamma=190$~keV)
 around the $\alpha+\alpha+t$ threshold with CSM.
The large radius ($R_{N}=5.98$~fm) indicates that the state has a dilute
 cluster structure.
The analysis of the single-cluster properties showed that this state
 has a main configuration of $(0S_{\alpha})^2(0S_t)$ with
 about $65$\% probability.
Thus, we could call the $1/2^{+}_{2}$ state the Hoyle-analogue with $(0S_{\alpha})^2(0S_t)$, 
 which is similar to the Hoyle state possessing
 the main configuration $(0S_{\alpha})^3$ with about $70$\% probability.
It should be reminded that $1/2^{+}_{2}$ is located by $0.75$ MeV above
 the $\alpha+\alpha+t$ threshold, while $1/2^{+}_{1}$ is bound by $4.2$ MeV
 with respect to the three cluster threshold.
The latter binding energy leads to the suppression of the development of
 the gas-like $\alpha+\alpha+t$ structure in $1/2^{+}_{1}$, whereas 
 the gas-like structure with a large nuclear radius grows up in $1/2^{+}_{2}$
 because the state appears above the three-body threshold.
 
Recently, the $1/2^{+}~(3/2^{+})$ state at $E_{x}=12.56$ MeV with $\Gamma=210\pm20$ keV
 (located at $1.4$ MeV above the $\alpha+\alpha+t$ threshold)
 was observed through the $\alpha$+$^7$Li decay channel~\cite{soic04,curtis05,charity08}.
The experimental energy and width of the $12.56$-MeV state are
 in good correspondence to the present calculated results of the $1/2^{+}_{2}$ state.
The Hoyle-analogue state in $^{11}$B, thus, could be assigned as 
 the $12.56$-MeV state.

\acknowledgments

The present authors thank to H.~Horiuchi, K.~Ikeda, G.~R\"opke, P.~Schuck, and A.~Tohaski for
 valuable discussions and encouragements.
This work was partially supported by JSPS (Japan Society for the Promotion of Science) Grant-in-Aid for
 Scientific Research (C) (21540283) and Young Scientists (B) (21740209).

\end{document}